\documentclass[aps,prd,showpacs,notitlepage,nofootinbib,superscriptaddress,floatfix,showkeys,twocolumn]{revtex4-1}
\usepackage{upgreek}
\usepackage{float}
\newcommand{\sumint}{
  \mathop{
    \mathchoice
      {\ooalign{$\displaystyle\sum$\cr\hidewidth$\displaystyle\int$\hidewidth\cr}}
      {\ooalign{$\textstyle\sum$\cr\hidewidth$\textstyle\int$\hidewidth\cr}}
      {\ooalign{$\scriptstyle\sum$\cr\hidewidth$\scriptstyle\int$\hidewidth\cr}}
      {\ooalign{$\scriptscriptstyle\sum$\cr\hidewidth$\scriptscriptstyle\int$\hidewidth\cr}}
  }
}
\usepackage{comment}
\usepackage{cancel}
\usepackage[normalem]{ulem} 
\usepackage{soul}
\usepackage{amsmath}
\usepackage{siunitx}
\usepackage{caption}
\usepackage{physics}
\usepackage{graphicx}
\usepackage{gensymb}
\usepackage{amsmath, cases}
\usepackage[usenames]{color}
\usepackage{amssymb}
\usepackage{hyperref}
\usepackage{wrapfig}
\usepackage{cleveref}
\usepackage{tikz}
\newcommand{\be}{\begin{equation}}
\newcommand{\ee}{\end{equation}}
\newcommand{\bea}{\begin{aligned}}
\newcommand{\eea}{\end{aligned}}
\newcommand{\pr}{\partial}

\newcommand{\bse}{\begin{subequations}}
\newcommand{\ese}{\end{subequations}}

\usepackage{bm}

\newcommand{\bmm}{\begin{multline}}
\newcommand{\emm}{\end{multline}}

\newcommand{\nt}{\notag\\}
\numberwithin{equation}{section}
\renewcommand{\thesection}{\Roman{section}}
\renewcommand{\theequation}{\arabic{section}.\arabic{equation}}
\begin{document}
\title{Canonical quantization of massive vector field in Schwarzschild black hole background}
\author{Chandra Prakash}
\email{chandra.pp@alumni.iitg.ac.in}
\affiliation{Department of Physics, 
Indian Institute of Technology Guwahati, Assam 781039, India}
\author{Rajesh Karmakar}
\email{rajesh@shu.edu.cn}
\affiliation{Department of Physics, Shanghai University, 99 Shangda Road, Shanghai, 200444, China}

%\maketit
\allowdisplaybreaks

\begin{abstract} 
We perform a first-principles canonical quantization of a massive vector field, often referred to as the Proca field, in a Schwarzschild spacetime background. While scalar, fermionic, and electromagnetic fields are well studied in this context, the Proca field requires a more nuanced treatment because of the physical nature of the longitudinal polarization mode and the constrained dynamics of the field variables. By implementing the Dirac bracket formalism to treat the constraints inherent in the Proca action, we derive a consistent framework for the commutator algebra of creation and annihilation operators. Following this construction, we define the usual Boulware, Unruh, and Hartle–Hawking vacua. Using the Unruh vacuum, we derive and analyze the Hawking spectrum of the Proca field. Furthermore, we numerically evaluate the Proca condensate constructed from the two-point correlation function $\langle A_\mu(x) A_\nu(x') \rangle$, defined on all three vacuum states. We find that the condensate becomes significant near the boundary of the future horizon. Our results highlight the interplay among the different polarization modes and the significance of the Proca mass in quantum observables.
\end{abstract}
%The present analysis provides the necessary foundation for studying the quantum backreaction and entanglement properties of dark photon candidates in the vicinity of black holes.
%distinctive behaviour of vector fluctuations near the event horizon, particularly for modes where the frequency $\omega$ is below the mass $\mu$, for which the field exhibits quasi-bound state characteristics. 
%%symplectic structure and a well-defined inner product for the field modes. Simply, this

\maketitle

\newpage
\section{Introduction}
The semiclassical approach to gravitational interactions, in which quantum fields propagate on a fixed classical spacetime background, captures several profound phenomena. It has led to successful predictions in early-universe cosmology, particularly regarding structure formation \cite{Kofman:1997yn}. On the other hand, the same approach led to the discovery of Hawking radiation \cite{Hawking:1975vcx} and vacuum polarization \cite{Candelas:1980zt} in BH spacetimes, opening an entirely new arena of phenomena associated with BH physics. While the implementation of quantization within the semiclassical framework is relatively well understood in the early universe \cite{Kolb:2023ydq}, where the background is isotropic, homogeneous, and expanding, difficulties arise in the presence of horizons in BH spacetimes (see, e.g., \cite{Birrell:1982ix, Parker:2009uva}). Consequently, substantial effort has been devoted over the decades to the quantization of fundamental fields in BH backgrounds and to the computation of vacuum expectation values of the stress-energy tensor \cite{Davies:1976ei, Unruh:1976db, Candelas:1984pg, Howard:1984qp}. 

Even decades after Hawking’s original discovery \cite{Hawking:1975vcx}, together with several alternative derivations of the radiation flux \cite{Hartle:1976tp, Israel:1976ur, Balbinot:1999ri, Srinivasan:1998ty, Parikh:1999mf, Robinson:2005pd, Majhi:2008gi, Banerjee:2009wb, Barman:2017fzh, Banerjee:2008ry}, a complete understanding of the semiclassical evaporation process remains an active area of research. Such an understanding requires the computation of the renormalized stress-energy tensor in order to study the semiclassical Einstein equation, $G_{\alpha\beta}=\langle T_{\alpha\beta}\rangle$ \cite{Christensen:1977jc}. As a first step, a major line of investigation has focused on the canonical quantization of fields. Naturally, the procedure becomes increasingly difficult as the spin of the matter field increases, as is evident from the existing literature on spin-0 (scalar) \cite{Balakumar:2022yvx, Egorov:2022hgg, Anderson:1993if}, spin-$1/2$ (fermion) \cite{Boulware:1975pe}, and massless spin-1 (photon) fields \cite{Volobuev:2024tyy, crispino2001quantization, Candelas:1981zv}. However, the inclusion of mass introduces additional complications. For example, anomaly-based methods can no longer be utilized to study near-horizon features \cite{Christensen:1977jc}. Nevertheless, within the canonical framework, the extension to the massive case is comparatively straightforward for scalar and fermionic fields, allowing the field modes to exhibit bound states in addition to scattering states \cite{Egorov:2022hgg}. In contrast, the corresponding extension for a vector field is nontrivial because the mass term breaks gauge invariance and thereby promotes the longitudinal polarization to a physical propagating degree of freedom. From a computational perspective, this results in a coupled system of equations that significantly complicates the numerical analysis \cite{Herdeiro:2011uu}. A major breakthrough came with the work of Frolov, Krtou\v{s}, Kubiz\v{n}'ak, and Santos (FKKS), who established an exact separable ansatz for the Proca equations in BH spacetimes \cite{Frolov:2018ezx}.

The motivation for a rigorous study of the Proca field (hereafter used interchangeably with massive vector field) in a BH background is twofold. First, from a phenomenological perspective, massive vector bosons, such as dark photons, constitute compelling dark matter candidates \cite{Kribs:2022gri,Fabbrichesi:2020wbt,Marriott-Best:2025sez,March-Russell:2022zll}. The longitudinal polarization of a massive vector field, which can be understood through the Stueckelberg mechanism \cite{Stueckelberg:1938hvi, Ruegg:2003ps, Belokogne:2015etf}, represents a physical degree of freedom that is absent in standard Maxwell electrodynamics. Understanding how quantization affects the dynamics and thermal properties of this additional mode is therefore of direct physical interest.

The explicit breaking of $U(1)$ gauge symmetry by the Proca mass term also raises an interesting question in the context of infrared approaches to the BH information problem. Proposals that relate information retention to soft hair and large gauge transformations at null infinity rely crucially on the existence of asymptotic gauge symmetries and their associated soft sectors~\cite{Hawking:2016msc}. In contrast, a massive vector field possesses no corresponding large $U(1)$ symmetry and therefore lacks the standard soft-photon hair structure familiar from massless electrodynamics. Furthermore, the study of massive vector fields in four dimensions has direct implications for string-inspired phenomenology through their duality with the massive Kalb-Ramond (KR) two-form field~\cite{Hell:2021wzm}. Since the quantization framework developed here is formulated at the free-field level, both the equations of motion and the canonical quantization procedure carry over directly onto the KR field under this duality.

In the present work, we study the massive vector field in the test-field approximation on a fixed Schwarzschild background, leaving the extension to rotating spacetimes for future investigation.
%Neglecting backreaction on the geometry, the Proca field, therefore, can be decomposed into a classical part and quantum fluctuations, $ \hat{A}_{\mu} = A_{\mu}^{\rm cl} + \delta\hat{A}_{\mu}$. We focus on the two-point function of the fluctuations, which encodes the leading quantum corrections to the stress-energy tensor.
We begin by constructing the classical solutions using the FKKS ansatz and explicitly verifying that the modes are orthonormal with respect to the Proca inner product. This step is particularly important in the even-parity sector, where the orthogonality of the longitudinal and transverse degrees of freedom is not guaranteed by parity alone. Using these orthonormal mode functions, we then perform a canonical quantization by following the Dirac--Bergmann approach \cite{Dirac:1950pj, Bergmann:1949zz, Bergmann:1954tc, Date:2010xr, das2020lectures} to handle the second-class constraints of the system \cite{dirac2013lectures} and establish the corresponding consistent bracket structure. Subsequently, we define the Boulware, Unruh, and Hartle--Hawking vacuum states.

As an important quantum observable, the number operator is evaluated in the past Unruh state to derive the Hawking spectrum at future null infinity. Compared to the massless case \cite{Page:1976df}, it is important to understand how the lifting of degeneracy affects the Hawking spectrum of the massive field. Motivated by this question, we analyze the spectrum associated with each polarization of the Proca field. On the other hand, as a fundamental building block for the quantum stress-energy tensor and the study of backreaction, the two-point Green function must be computed as a first step \cite{Birrell:1982ix, Parker:2009uva}. Accordingly, we evaluate this quantity in all three vacuum states and subsequently investigate the structure of the Proca condensate through numerical analysis. 

The rest of the paper is organized as follows. In \Cref{sec: classical_soln}, we use the FKKS ansatz to derive the classical solutions of the EoM, which form the foundation of our analysis. In \Cref{sec: asymptoticform}, we impose the appropriate boundary conditions on these solutions and establish the key properties and identities satisfied by the reflection and transmission coefficients. Building on these results, in \Cref{sec: innerproduct and normalization} we introduce the Proca inner product and construct an orthonormal basis of classical solutions. Using this basis, in \Cref{sec: quantization_and_vacua} we 
%promote the Dirac brackets to commutators and 
carry out the canonical quantization procedure, consistently deriving the commutation relations for the creation and annihilation operators from those of the field and its conjugate momenta. We then perform the necessary analytic continuations and define the relevant vacuum states. In \Cref{sec: hawkingflux}, we derive the Hawking flux from first principles. Next, in \Cref{sec: vevs}, we compute the two-point function and discuss its physical implications. Finally, in \Cref{sec: conclusion}, we summarize our results and outline possible directions for future work.

In the following analysis, we work in units where $\hbar = c = 1$.
%set the horizon radius to unity, $r_h = 1$.
%%%%%%%%%%%%%%%%%%%%%%%%%%%%%%%%%%%%%%%%%%%%%%%%%%%%%%%%%%%%%%%%%%%%%%%%%%%
\section{Classical solutions of the massive vector field in Schwarzschild spacetime}\label{sec: classical_soln}
As a first case study of the Proca field, we consider a static, spherically symmetric BH spacetime described by the Schwarzschild line element,
\be\label{eq: schw_metric}
ds^2=-f(r)dt^2+f(r)^{-1}dr^2+r^2d\theta^2+r^2\sin^2\theta d\varphi^2,
\ee
where $f(r)=1-r_h/r$, with $r_h=2GM$, and $M$ denotes the BH mass. In a generic curved spacetime, the action for a Proca field minimally coupled to the background geometry \eqref{eq: schw_metric} is given by
\be\label{eq: proca_action}
S=-\int d^4x\sqrt{-g}\left(\frac14 F_{\mu\nu}F^{\mu\nu}+\frac12 \mu^2 A_\mu A^\mu\right),
\ee
where $\mu$ denotes the Proca mass. The corresponding equations of motion are
\be\label{eq: eom1}
\frac{1}{\sqrt{-g}}\pr_\mu\left(\sqrt{-g}F^{\mu\nu}\right)=\mu^2 A^\nu.
\ee
In a Ricci-flat spacetime, it follows from the above field equations that
\be\label{eq: gauge.cond}
\pr_\mu\left(\sqrt{-g}A^\mu\right)=0.
\ee
This constraint, commonly referred to as the Lorenz condition \cite{Rosa:2011my}, serves as an integrability condition for the Proca equations. Unlike in Maxwell electrodynamics, where the Lorenz condition is imposed as a gauge choice, here it arises directly from the field equations and therefore need not be imposed separately on the field modes. Before proceeding to quantization, however, one must first construct the classical solutions that define the mode basis. Accordingly, in the following discussion, we outline the treatment of the Proca equations, beginning with a suitable decomposition that allows the equations of motion to be decoupled.
%%%%%%%%%%%%%%%%%%%%%%%%%%%%%%%%%%
\subsection{Vector spherical harmonic (VSH) basis}
%%%%%%%%%%%%%%%%%%%%%%%%%%%%%%%%
For a spin--$1$ field in the spherically symmetric spacetime \eqref{eq: schw_metric}, it is convenient to describe the angular sector using the following set of vector spherical harmonics (VSH) \cite{Rosa:2011my}:

\be
\bea
&Z^{(1)\ell m}_\mu=\left[1,0,0,0\right]Y^{\ell m}(\Omega),\\
&Z^{(2)\ell m}_\mu=\left[0,f^{-1},0,0\right]Y^{\ell m}(\Omega),\\
&Z^{(3)\ell m}_\mu=\frac{r}{\sqrt{L}}\left[0,0,\pr_\theta,\pr_\phi\right]Y^{\ell m}(\Omega),\\
&Z^{(4)\ell m}_\mu=\frac{r}{\sqrt{L}}\left[0,0,\frac{1}{\sin\theta}\pr_\phi,-\sin\theta\pr_\theta\right]Y^{\ell m}(\Omega),\\
\eea
\ee
where $L=\ell(\ell+1)$, and $Y_{\ell m}(\Omega)$, with $\Omega\equiv (\theta,\varphi)$, denote the usual scalar spherical harmonics. The above expressions satisfy the following orthogonality condition

\be\label{eq: vsh_cond}
\int\left(Z^{(i)\ell m}_\mu\right)^*\eta^{\mu\nu}\left(Z^{(i')l'm'}_\nu\right)\sin\theta d\theta d\varphi=\delta_{ii'}\delta_{\ell\ell'}\delta_{mm'},
\ee
with 
\be
\eta_{\mu\nu}={\rm diag}[1,f^2,1/r^2,1/(r^2\sin^2\theta)].
\ee
Concerning the transformation under parity, $(\theta,\varphi)\to (\pi-\theta, \varphi+\pi)$, the VSH basis exhibits the following properties:

\be
\bea
&Z^{(1,2,3)\ell m}=(-1)^lZ^{(1,2,3)\ell m},\\
&Z^{(4)\ell m}=(-1)^{(l+1)}Z^{(4)\ell m},
\eea
\ee
corresponding to even and odd parity, respectively. Since $Z_{\mu}$ are parity eigenstates, it enforces that the even parity sector is orthogonal to the odd parity sector. With this setup, we decompose the vector field as

\be\label{eq: proca_decomp}
A_\mu(t,{\bf r})=\frac{1}{r}\sum^4_{i=1}\sum_{\ell m}c_{(i)}u_{(i)}^{\ell}(t,r)Z^{(i)\ell m}_\mu(r, \Omega),
\ee
with $c_{(1)}=c_{(2)}=1$ and $c_{(3)}=c_{(4)}=1/\sqrt{L}$, whereas (${\bf r}$) stands for ($r,\theta,\phi$). Substituting the above decomposition into \eqref{eq: eom1},  equations of motion for different parity sectors with decoupled angular part can be obtained. This yields a set of coupled equations in the even-parity sector \cite{Hancock:2025ois},

\be\label{eq: coupled_eqn}
\bea
&\hat{\mathcal{D}}_2u_{1}+\left[(\pr_{r}f)\left(\pr_t{u}_{(2)}-\pr_{r_*}u_{(1)}\right)\right]=0,\\
&\hat{\mathcal{D}}_2u_{2}+\left[(\pr_{r}f)\left(\pr_t{u}_{(1)}-\pr_{r_*}u_{(2)}\right)-\frac{2f^2}{r^2}\left(u_{(2)}-u_{(3)}\right)\right]=0,\\
&\hat{\mathcal{D}}_2u_{3}+\left[\frac{2}{r^2}f\ell(\ell+1)u_{(2)}\right]=0,\\
\eea
\ee
with 
\be
\hat{\mathcal{D}}_2\equiv -\pr^2_t+\pr^2_{r_*}-f\left[\frac{L}{r^2}+\mu^2\right],
\ee
where $r_*$ denotes the Tortoise coordinate, defined as $dr_*=f^{-1}dr$. The special case $l=0$, corresponding to the monopole mode (${\rm m}$), belongs to the even-parity sector and is governed by the following equation:

\be\label{de:mono}
-\pr^2_tu_{(2)}+\pr^2_{r_*}u_{(2)}-V^{\rm (m)}_{\rm eff}(r)u_{(2)}=0,
\ee
with 
\be
V^{\rm (m)}_{\rm eff}(r)=f\left(\frac{2f}{r^2}-\frac{f'}{r}+\mu^2\right).
\ee
Given the static nature of Schwarzschild spacetime, it is suitable to consider the following separable form:

\be\label{eq: ansatz_u}
u_{(i)}(t,r)=e^{-i\omega t}u_{(i)}(r, \omega).
\ee
Implementing the above decomposition, the equation for the monopole mode \eqref{de:mono} takes the following form
\be\label{eq: monopole_eom}
\pr^2_{r_*}u_{(2)}+\left[\omega^2-V^{\rm (m)}_{\rm eff}(r)\right]u_{(2)}=0.
\ee
For non-zero $\ell$ modes, in addition to the even parity equations above, we obtain a fully decoupled equation governing the odd-parity sector,
\be
\mathcal{D}_2u_{(4)}=0.
\ee
Separating the time dependence in the same manner as in \eqref{eq: ansatz_u}, the above equation reads 
\be\label{eq: odd_parity_eom}
\pr^2_{r_*}u_{(4)}+\left[\omega^2-V^{\rm ({\rm o})}_{\rm eff}(r)\right]u_{(4)}=0,
\ee
with 
\be
V^{\rm ({\rm o})}_{\rm eff}(r)=f\left(\frac{L}{r^2}+\mu^2\right).
\ee
Therefore, we are left with the set of coupled equation for $l \neq 0$ \eqref{eq: coupled_eqn} in the even-parity sector. To simplify the construction of solutions, it is convenient to decouple these equations, and the corresponding procedure is discussed next.
%%%%%%%%%%%%%%%%%%%%%%%%%%%%%%%%%
\subsection{Frolov--Krtou\v{s}--Kubiz\v{n}\'{a}k--Santos (FKKS) basis}
%%%%%%%%%%%%%%%%%%%%%%%%%%%%%%%%%%%%
In the even-parity sector \eqref{eq: coupled_eqn}, the equations of motion do not admit a straightforward decoupling. To address this, we adopt the Frolov–Krtouš–Kubizňák–Santos (FKKS) approach \cite{Frolov:2018ezx}, previously implemented in \cite{Percival:2020skc, Fernandes:2021qvr}. The strategy is to rewrite the components governing the radial part in the even parity sector as follows:
\be\label{fkks:decomp}
\bea
&u_{(1)}(\omega, r)=-\frac{ifr(\nu r\pr_r +\omega/f)}{q_r}\bar{R}(\omega, r),\\
&u_{(2)}(\omega, r)=\frac{fr(\pr_r-\omega\nu r/f)}{q_r}\bar{R}(\omega, r),\\
&u_{(3)}(\omega, r)=L\bar{R}(\omega, r),
\eea
\ee
with $q_r = 1+\nu^2 r^2$. Where $\nu$ is a separation constant, to be determined below. Substituting the expression for $u_{(2)}$ and $u_{(3)}$ into the last equation of the set \eqref{eq: coupled_eqn}, the governing equation of $\bar{R}$ can be obtained straightforwardly and written as
\be\label{eq: Rbar_eqn}
\pr^2_{r_*}+\frac{2f}{rq_r}\pr_{r_*}+\left[\omega^2-f\left(\frac{L}{r^2}+\mu^2+\frac{2\omega\nu}{q_r}\right)\right]\bar{R}=0.
\ee
To determine the separation constant, $\nu$, we substitute the components from \eqref{fkks:decomp} into the Lorenz condition \eqref{eq: gauge.cond}, which yields
\be
\pr^2_{r_*}+\frac{2f}{rq_r}\pr_{r_*}+\left[\omega^2-f\left(3\omega\nu+\frac{q_rL}{r^2}-\frac{2\omega\nu^3r^2}{q_r}\right)\right]\bar{R}=0.
\ee
The equivalence of the preceding two equations implies the relation
\be
L\nu^2+\omega \nu-\mu^2=0.\label{eq22}
\ee
The roots of this equation are given by
\be
\nu_\pm=-\frac{\omega}{L}\left[\frac{1\pm \sqrt{1+4L\mu^2/\omega^2}}{2}\right].
\ee
For the monopole mode, we discard the $\nu_{+}$ branch and retain $\nu_{-}=\mu^2/\omega$, in which case the resulting equation reduces to \eqref{de:mono}. Importantly, the FKKS ansatz remains well defined in the massless limit. In this limit, the $\nu_{-}$ branch continuously reduces to a pure gauge configuration ~\cite{Krtous:2018bvk}. Since the longitudinal polarization resides in the even parity sector, taking $\mu\to 0$ implies $\nu_{-}\to 0$, and the associated vector field reduces to the gradient of a scalar function. On the other hand, the positive branch,  $\nu_+$ in the massless limit corresponds to the transverse vector mode as in the case of the photon.  For these reasons, the $(+)/(-)$ labels in the even parity sector correspond to vector- and scalar-type modes, respectively \cite{Hancock:2025ois}.
\begin{comment}
%This is expected because the FKKS construction automatically satisfies the Lorenz condition, thereby selecting a particular gauge representative rather than preserving manifest gauge invariance .
It is important to emphasize that the Lorenz condition alone does not completely fix the gauge. Residual gauge transformations satisfying the homogeneous wave equation still remain, reflecting the well-known Gribov ambiguity~\cite{Sainapha:2019cfn}. Consequently, an additional gauge-fixing condition is required to eliminate the residual pure gauge degree of freedom in the massless theory.
\end{comment}

To express the governing equations for both even and odd parity modes in a unified Schr\"odinger-like form, we perform the following rescaling $\bar{R}\to \bar{R}\sqrt{q_r}/r$. With this choice, the governing equation for $\bar R(r)$ in the even-parity sector becomes
\be\label{eq: even_parity_eom}
\left[\pr^2_{r_*}+\omega^2-V^{\rm even(\pm)}_{\rm eff}\right]\bar{R}=0,
\ee
where the effective potential is given by
\be
V^{\rm even(\pm)}_{\rm eff}=V^{\rm odd}_{\rm eff}+\pr_{r_*}\left(\frac{f}{rq^{(\pm)}_r}\right)+\left(\frac{f}{rq^{(\pm)}_r}\right)^2+\frac{2f\omega \nu_\pm}{q^{(\pm)}_r}.
\ee
\begin{comment}
Having obtained the decoupled equations of motion in the form  \eqref{eq: monopole_eom}, \eqref{eq: odd_parity_eom}, and \eqref{eq: even_parity_eom}, we now proceed to discuss the corresponding solutions. The equations of motions outlined above have been studied previously; and in particular, analytical solutions for the monopole and odd parity sectors have been obtained in various regimes of the field mass in Ref.~\cite{Hancock:2025ois}. It has been shown that these solutions can be expressed in terms of confluent Heun functions. 
%Predictably, similar Heun-type structures also arise in the radial equations for massive scalar fields in Schwarzschild~\cite{philipp2015analytic} and Kerr--Newman spacetimes~\cite{Bezerra:2013iha, Vieira:2014waa}.
On the other hand, because of the complicated form of the frequency-dependent effective potential in the even parity sector, the corresponding solutions have been investigated numerically \cite{Hancock:2025ois}. In the present work, however, we follow a fully numerical procedure, as recently studied in \cite{Vispute:2026vek}, leaving the analytical estimate for future work. 
\end{comment}
An interesting property of the underlying framework, as can be seen from all the linear second-order master equations, is that they admit a conserved Wronskian  \cite{unruh1976absorption}, which can be expressed using \eqref{eq: even_parity_eom} as,
\be
W(\bar{R}_1, \bar{R}_2) =\frac{r^2}{q_r}(\bar R_1^*\pr_{r_*}\bar R_2-\pr_{r_*}\bar R^*_1\, \bar R_2),\label{conserved:wronskian}
\ee
where $\bar R_1$ and $\bar R_2$ are two linearly independent solutions of \eqref{eq: even_parity_eom}. This conserved quantity, as we will see, represents the radial flux associated with the mode function. The presence of the nontrivial measure $r^2/q_r$ appears because of the rescaling $\bar{R}\to \bar{R}\sqrt{q_r}/r$ in \eqref{eq: even_parity_eom}.
Similarly, the odd parity sector \eqref{eq: odd_parity_eom} also has an associated conserved Wronskian, which has a similar form as above, excluding the overall factor.

Having obtained the decoupled equations of motion in the form \eqref{eq: monopole_eom}, \eqref{eq: odd_parity_eom}, and \eqref{eq: even_parity_eom}, one can straightaway impose the relevant boundary conditions and thereby construct an orthonormal basis from the classical solutions. Before proceeding, however, for convenience in the analysis, we find it useful to express the Proca field in a suitable form using the combination of VSH and FKKS bases, as discussed below.
%%%%%%%%%%%%%%%%%%%%%%%%%%%%%%%%%%
\subsection{Combined VSH and FKKS basis:}
%%%%%%%%%%%%%%%%%%%%%%%%%%%%%%%%%%
From the above discussion, it becomes clear that the FKKS decomposition resolves the difficulties associated with decoupling of the equation of motion in the even parity sector for $\ell>0$. In addition, for $\ell=0$, it captures the monopole mode associated with the scalar-type branch ($\nu_{-}$), which is already decoupled in the VSH basis. The relevance of both bases has also been emphasized in studies of the Proca field over the years, where monopole dynamics in the even parity sector plays a distinguished role in the structure of the perturbation equations, alongside the odd parity modes~\cite{Ge:2025yqk, Konoplya:2005hr, Konoplya:2006gq}. This suggests that an effective decomposition of the Proca field should incorporate both the parity sectors. We find that this can be achieved by combining the FKKS and VSH constructions in the following way:
\begin{widetext}
\be
\bea
A_{t}(t,\mathbf{r})&=e^{-i\omega t}\sum_{\ell m}\left(-\frac{i\omega}{q_r}-\frac{i\nu_\pm r f}{q_r}\,\partial_r\right)\mathcal{N}^{\rm(e\pm)}_{\omega \ell }\bar{R}_{\omega \ell}^{(\rm e\pm)}(r) Y^{\ell m}(\Omega),\\
A_{r}(t,\mathbf{r})&=e^{-i\omega t}\sum_{\ell m}\left(\frac{1}{q_r}\,\partial_r- \frac{\omega \nu_\pm r}{q_r f}\right)\mathcal{N}^{\rm(e\pm)}_{\omega \ell }\bar{R}_{\omega \ell}^{\rm (e\pm)}(r) Y^{\ell m}(\Omega),\\
A_{\theta}(t,\mathbf{r})&=e^{-i\omega t}\sum_{\ell m}\left[\mathcal{N}^{\rm(e\pm)}_{\omega \ell }\bar{R}_{\omega \ell}^{\rm (e\pm)}(r)\pr_\theta Y^{\ell m}(\Omega)+\mathcal{N}^{\rm({\rm o})}_{\omega \ell }\frac{\bar{R}_{\omega \ell}^{\rm ({\rm o})}(r)}{L}\frac{\pr_\varphi Y^{\ell m}(\Omega)}{\sin\theta}\right],\\
A_{\varphi}(t,\mathbf{r})&=e^{-i\omega t}\sum_{\ell m}\left[\mathcal{N}^{\rm(e\pm)}_{\omega \ell }\bar{R}_{\omega \ell}^{\rm (e\pm)}(r)\pr_\phi Y^{\ell m}(\Omega)-\mathcal{N}^{\rm({\rm o})}_{\omega \ell }\frac{\bar{R}_{\omega \ell}^{\rm ({\rm o})}(r)}{L}\sin\theta\pr_\theta Y^{\ell m}(\Omega)\right],
\eea
\ee
\end{widetext}
In the above decomposition, we have also included the overall normalization factors, $\mathcal{N}^{\rm(e\pm)}_{\omega \ell }$ and $\mathcal{N}^{\rm({\rm o})}_{\omega \ell }$ for even and odd parity, respectively, which will be fixed in \Cref{sec: innerproduct and normalization}. 
%Nevertheless, from the above decomposition, the odd and even parity sectors can be realized.
For convenience in the quantization procedure, we label Proca modes in the respective parity sector:
\begin{widetext}
{\it Even parity sector:}

\be\label{eq: combined_even_basis}
A^{(\rm e\pm)}_\mu(t,\mathbf r)=e^{-i\omega t}\sum_{\ell m}\mathcal{N}^{\rm(e\pm)}_{\omega \ell }
\begin{pmatrix}
\left(-\frac{i\omega}{q_r}-\frac{i\nu_\pm r f}{q_r}\partial_r\right)\bar{R}^{(\rm e\pm)}_{\omega \ell}(r)\, Y^{\ell m}(\Omega)\\
\left(\frac{1}{q_r}\,\partial_r- \frac{\omega \nu_\pm r}{q_r f}\right)\bar{R}_{\omega \ell}^{(\rm e\pm)}(r)\, Y^{\ell m}(\Omega)\\
\bar{R}^{(\rm e\pm)}_{\omega \ell}(r)\,\partial_\theta Y^{\ell m}(\Omega)\\
\bar{R}^{(\rm e\pm)}_{\omega \ell}(r)\,\partial_\varphi Y^{\ell m}(\Omega)
\end{pmatrix}.
\ee
{\it Odd parity sector:}
\be\label{eq: combined_odd_basis}
A^{\rm ({\rm o})}_\mu(t,\mathbf r)
=e^{-i\omega t}\sum_{\ell m}\mathcal{N}^{\rm({\rm o})}_{\omega \ell }\frac{\bar{R}^{\rm ({\rm o})}_{\omega \ell}(r)}{L}
\begin{pmatrix}
0\\
0\\
\frac{\partial_\varphi Y^{\ell m}(\Omega)}{\sin\theta}\\
-\sin\theta\,\partial_\theta Y^{\ell m}(\Omega)
\end{pmatrix}.
\ee
\end{widetext}
In the rest of the analysis, the combined set of modes, altogether, will be referred to as $A^{(\zeta)}_\mu$, with $\zeta\equiv \lambda n; \omega \ell  m$, where $\lambda$ takes care of the parity sectors $(\rm e\pm)$ and $(\rm o)$, whereas $n$ stands for the ${\rm in, up, out}$ and ${\rm down}$ modes, which will be discussed in the following section.  
%%%%%%%%%%%%%%%%%%%%%%%%%%%%%%%%%%%%%%%%%%%
\section{Asymptotic forms of the Proca field}\label{sec: asymptoticform}
\begin{figure}[h]
    \centering
    \resizebox{1\linewidth}{!}{
    \begin{tikzpicture}[
        every node/.style={font=\fontsize{18}{20}\selectfont}
    ]

    \newcommand{\DiamondWithGrid}[3]{
        \begin{scope}[shift={(#1,#2)}]
            \begin{scope}
                \clip (0,-#3) -- (#3,0) -- (0,#3) -- (-#3,0) -- cycle;
                \begin{scope}[rotate=45]
                    \draw[gray!30, step=0.5] (-10,-10) grid (10,10);
                \end{scope}
            \end{scope}
            \draw[thick] (0,-#3) -- (#3,0) -- (0,#3) -- (-#3,0) -- cycle;
        \end{scope}
    }

    %-----------------------------------
    % IN MODES
    %-----------------------------------
    \DiamondWithGrid{0}{13}{5}

    \draw[blue, thick, ->, >=stealth]
        (2.5,10.5)--(-2.5,15.5);

    \draw[blue, thick, ->, >=stealth]
        (0,13)--(2.5,15.5);

    \node[right] at (5,13) {$i^0$};
    \node[above] at (0,18) {$i^+$};
    \node[below] at (0,8) {$i^-$};

    \node at (3,10) {$\mathcal{I}^-$};
    \node at (-3,10) {$\mathcal{H}^-$};
    \node at (-3,16) {$\mathcal{H}^+$};
    \node at (3,16) {$\mathcal{I}^+$};

    \node at (-3,18) {\textbf{in}};

    \draw[red, thick, ->, >=stealth]
        (1.775,10.52)--(-2.5,14.8);

    \draw[red, thick, ->, >=stealth]
        (1.75,11.6) arc (225:135:2);

    %-----------------------------------
    % UP MODES
    %-----------------------------------
    \DiamondWithGrid{13}{13}{5}

    \draw[blue, thick, ->, >=stealth]
        (10.5,10.5)--(15.5,15.5);

    \draw[blue, thick, ->, >=stealth]
        (13,13)--(10.5,15.5);

    \node[right] at (18,13) {$i^0$};
    \node[above] at (13,18) {$i^+$};
    \node[below] at (13,8) {$i^-$};

    \node at (16,10) {$\mathcal{I}^-$};
    \node at (10,10) {$\mathcal{H}^-$};
    \node at (10,16) {$\mathcal{H}^+$};
    \node at (16,16) {$\mathcal{I}^+$};

    \node at (10,18) {\textbf{up}};

    \draw[red, thick, ->, >=stealth]
        (11.23,10.52)--(15.5,14.8);

    \draw[red, thick, ->, >=stealth]
        (11.25,11.6) arc (-45:45:2);

    \end{tikzpicture}
    }
\caption{Representation of in- and up- radial modes originating from two different boundary conditions on the Penrose diagram of region I.}
    \label{fig:in-up}
\end{figure}
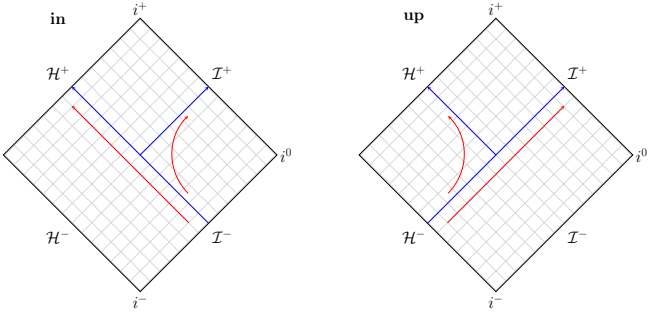

\begin{figure}[h]
    \centering
    \resizebox{1\linewidth}{!}{
    \begin{tikzpicture}[
        every node/.style={font=\fontsize{18}{20}\selectfont}
    ]

        \newcommand{\DiamondWithGrid}[3]{
            \begin{scope}[shift={(#1,#2)}]
                \begin{scope}
                    \clip (0,-#3) -- (#3,0) -- (0,#3) -- (-#3,0) -- cycle;
                    \begin{scope}[rotate=45]
                        \draw[gray!30, step=0.5] (-10,-10) grid (10,10);
                    \end{scope}
                \end{scope}
                \draw[thick] (0,-#3) -- (#3,0) -- (0,#3) -- (-#3,0) -- cycle;
            \end{scope}
        }

        %-----------------------------------
        % OUT MODES
        %-----------------------------------
        \DiamondWithGrid{0}{0}{5}

        \draw[thick] 
            (0,-5) -- (5,0) -- (0,5) -- (-5,0) -- cycle;

        \draw[blue, thick] 
            (0,0)--(2.5,-2.5);

        \draw[blue, thick, ->, >=stealth] 
            (-2.5,-2.5)--(2.5,2.5);

        \node[right] at (5,0) {$i^0$};
        \node[above] at (0,5) {$i^+$};
        \node[below] at (0,-5) {$i^-$};

        \node at (3,-3) {$\mathcal{I}^-$};
        \node at (-3,-3) {$\mathcal{H}^-$};
        \node at (-3,3) {$\mathcal{H}^+$};
        \node at (3,3) {$\mathcal{I}^+$};

        \node at (-3,5) {\textbf{out}};

        \draw[red, thick, ->, >=stealth]
            (-2.475,-1.775)--(1.79,2.5);

        \draw[red, thick, <-, >=stealth]
            (1.75,1.5) arc (135:225:2);

        %-----------------------------------
        % DOWN MODES
        %-----------------------------------
        \DiamondWithGrid{13}{0}{5}

        \draw[blue, thick, ->, >=stealth]
            (15.5,-2.5)--(10.5,2.5);

        \draw[blue, thick]
            (13,0)--(10.5,-2.5);

        \node[right] at (18,0) {$i^0$};
        \node[above] at (13,5) {$i^+$};
        \node[below] at (13,-5) {$i^-$};

        \node at (16,-3) {$\mathcal{I}^-$};
        \node at (10,-3) {$\mathcal{H}^-$};
        \node at (10,3) {$\mathcal{H}^+$};
        \node at (16,3) {$\mathcal{I}^+$};

        \node at (10,5) {\textbf{down}};

        \draw[red, thick, ->, >=stealth]
            (15.47,-1.78)--(11.22,2.5);

        \draw[red, thick, <-, >=stealth]
            (11.25,1.4) arc (45:-45:2);

    \end{tikzpicture}
    }
    
    \caption{Representation of out and down radial modes on the Penrose diagram of region I.}
    
    \label{fig:out-down}
\end{figure}
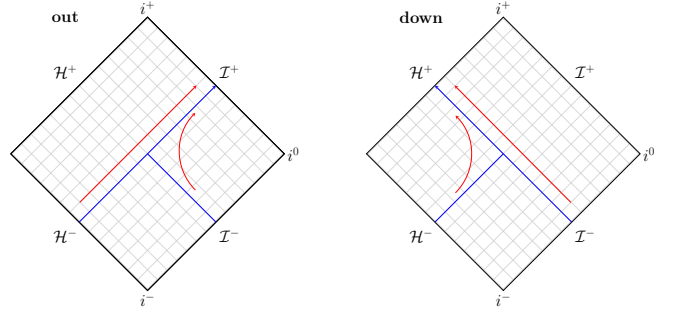

In this section, we discuss the asymptotic structure of the Proca field, which will be used to construct an orthonormal basis of mode functions in later sections. In the near-horizon limit, the equation governing the $(\rm e\pm)$ modes, \eqref{eq: even_parity_eom}, can be expressed as
\begin{align}
\left[\pr^2_{r_*}+\omega^2\right]\bar{R}^{\rm (e\pm)}_{\rm up}(\omega,r)=0,
\end{align}
whereas, the same equation \eqref{eq: even_parity_eom} in the limit $r \to \infty$ takes the following form:

\begin{align}
\left[\pr^2_{r}+\omega^2-\mu^2\right]\bar{R}^{\rm (e\pm)}_{\rm in}(\omega,r)=0.
\end{align}
The general solution to the above asymptotic equation can be expressed as the superposition of plane waves. In general, two different types of boundary condition define the in- and up-radial modes, as illustrated in Region-I of Fig.~\ref{fig:in-up}, for all three polarizations. Specifically, the modes incoming from the past null infinity are labeled as in-modes, whereas the modes outgoing from the past horizon are referred to as up-modes \cite{Balakumar:2022yvx}. Explicitly, these modes are characterized by the following asymptotic forms \cite{Rosa:2011my},

\begin{align}\label{Rin:BD}
\bar{R}^{(\lambda)\,\text{in}}_{\omega \ell}(r) &=
\begin{cases}
\mathcal{T}^{(\lambda)\,\text{in}}_{\omega \ell} e^{-i \omega r_*}, & r_* \to -\infty, \\
e^{-ik r_*} + \mathcal{R}^{(\lambda)\,\text{in}}_{\omega \ell } e^{i k r_*}, & r_* \to \infty,
\end{cases}
\end{align}
and

\begin{align}
\bar{R}^{(\lambda)\,\text{up}}_{\omega \ell}(r) &=
\begin{cases}
e^{i \omega r_*} + \mathcal{R}^{(\lambda)\,\text{up}}_{\omega \ell} e^{-i \omega r_*}, & r_* \to -\infty, \\
\mathcal{T}^{(\lambda)\,\text{up}}_{\omega \ell} e^{i k r_*}, & r_* \to \infty,
\end{cases}\label{Rup:BD}
\end{align}
With $k:=\sqrt{\omega^2-\mu^2}$, the above set of asymptotic solutions describes scattering states for $\omega>\mu$. For $\omega < \mu$, the relevant bound states correspond to up-modes that describe an outgoing wave at $\mathcal{H}^{-}$ and an ingoing wave at $\mathcal{H}^{+}$, with no flux reaching $\mathcal{I}^{+}$. We emphasize that these solutions are defined for $\omega \in \mathbb{R}$ and therefore do not correspond to quasi-bound states with a discrete spectrum, which require $\omega \in \mathbb{C}$ (see, e.g.,~\cite{Rosa:2011my}).
\begin{comment}
    In the null coordinate $ u = t-r_*;\quad v=t+r_*$
\begin{widetext}
\begin{align}
\bar{R}^{(\lambda)\,\text{in}}_{\omega \ell m}(r)e^{-i\omega t} &=
\begin{cases}
\mathcal{T}^{(\lambda)\,\text{in}}_{\omega \ell m} e^{-i \omega v}, & r_* \to -\infty, \\
e^{-iKv-ipu} + \mathcal{R}^{(\lambda)\,\text{in}}_{\omega \ell m} e^{-iKv-ipu}, & r_* \to \infty,
\end{cases}
\\[1em]
\bar{R}^{(\lambda)\,\text{up}}_{\omega \ell m}(r)e^{-i\omega t} &=
\begin{cases}
e^{-i \omega u} + \mathcal{R}^{(\lambda)\,\text{up}}_{\omega \ell m} e^{-i \omega v}, & r_* \to -\infty, \\
\mathcal{T}^{(\lambda)\,\text{up}}_{\omega \ell m} e^{-i K u-ipv}, & r_* \to \infty,
\end{cases}
\end{align}
\end{widetext}
\end{comment}
Nevertheless, in the next section, we will see that the in- and up-modes are orthogonal. By complex conjugation to these modes, we can also define the conjugate basis, out- and down-modes, as \cite{Balakumar:2022zyx},

\begin{align}
\bar{R}^{(\lambda)\,\text{out}}_{\omega \ell}(r) &= \bar{R}^{(\lambda)\,\text{in}*}_{\omega\ell}(r), &
\bar{R}^{(\lambda)\,\text{down}}_{\omega \ell}(r) &= \bar{R}^{(\lambda)\,\text{up}*}_{ \omega\ell}(r).
\end{align}
Consequently, the asymptotic forms can be expressed as,

\begin{align}
\bar{R}^{(\lambda)\,\text{out}}_{\omega \ell}(r) &=
\begin{cases}
    \mathcal{T}^{(\lambda)\,\text{in}*}_{\omega \ell} e^{i \omega r_*}, & r_* \to -\infty, \\
    e^{i k r_*} + \mathcal{R}^{(\lambda)\,\text{in}*}_{\omega \ell} e^{-ik r_*}, & r_* \to \infty,
\end{cases}\\[1em]
\bar{R}^{(\lambda)\,\text{down}}_{\omega \ell}(r) &=
\begin{cases}
    e^{-i \omega r_*} + \mathcal{R}^{(\lambda)\,\text{up}*}_{\omega \ell} e^{i \omega r_*}, & r_* \to -\infty, \\
    \mathcal{T}^{(\lambda)\,\text{up}*}_{\omega \ell} e^{-i k r_*}, & r_* \to \infty,
\end{cases}
\end{align} 
and illustrated in Fig.~\ref{fig:out-down}. Having given the asymptotic forms, the next task is to determine the reflection and transmission coefficients and their relations. A straightforward way to proceed is to utilize the fact that the propagation of modes, as shown in Fig.~\ref{fig:in-up} and Fig.~\ref{fig:out-down}, has associated conserved flux, because of which the induced inner product (defined in the next section) satisfies the following properties, connecting the past future null boundaries,
%%%%%%%%%%%%%%%

\begin{comment}
\begin{align}
    \begin{pmatrix}
        \bar{R}^{(e+)\rm in}_{\omega \ell m}\\
        \bar{R}^{(e-)\rm in}_{\omega \ell m}\\
        \bar{R}^{({\rm o})\rm in}_{\omega \ell m}\\
    \end{pmatrix}&=
    \begin{pmatrix}
        \mathcal{N}_{\omega \ell }^{(e+)\,\rm in}\\
        \mathcal{N}_{\omega \ell }^{(e-)\,\rm in}\\
        \mathcal{N}_{\omega \ell }^{({\rm o})\,\rm in}\\
    \end{pmatrix}e^{-ikr_{*}}\nt
    &+
    \begin{pmatrix}
        \mathcal{R}^{(e+)\, \rm in}_{\omega \ell m} & 0 &0\\
        0 & \mathcal{R}^{(e-)\, \rm in}_{\omega \ell m} &0\\
        0 & 0 &\mathcal{R}^{({\rm o})\, \rm in}_{\omega \ell m}
    \end{pmatrix}
    \begin{pmatrix}
        \mathcal{N}_{\omega \ell }^{(e+)\,\rm in}\\
        \mathcal{N}_{\omega \ell }^{(e-)\,\rm in}\\
       \mathcal{N}_{\omega \ell }^{({\rm o})\,\rm in}\\
    \end{pmatrix}e^{ikr_{*}}
\end{align}
\end{comment}
\begin{align}
    (A^{\rm in}_{\omega \ell m},A^{\rm in}_{\omega \ell m})_{\mathcal{I}^{-}}&= (A^{\rm in}_{\omega \ell m},A^{\rm in}_{\omega \ell m})_{\mathcal{H}^{+}}+ (A^{\rm in}_{\omega \ell m},A^{\rm in}_{\omega \ell m})_{\mathcal{I}^{+}},\\
    (A^{\rm up}_{\omega \ell m},A^{\rm up}_{\omega \ell m})_{\mathcal{H}^{-}}&= (A^{\rm up}_{\omega \ell m},A^{\rm up}_{\omega \ell m})_{\mathcal{H}^{+}}+ (A^{\rm up}_{\omega \ell m},A^{\rm up}_{\omega \ell m})_{\mathcal{I}^{+}}.
\end{align}
%Additionally using the (anti-)linearity of the inner product (see \ref{sec: innerproduct and normalization}), $
%(\Lambda A, A' )=\Lambda^*(A, A')$ and $(A,\Lambda A' )=\Lambda(A, A' )$, in the above expressions,

Utilizing these relations, one can show that the coefficients satisfy
\begin{align}\label{ref-trans-normalization}
    |\mathcal{R}^{(\lambda)\,\text{in}}_{\omega \ell}|^2+\frac{|\mathcal{N}^{(\lambda)\,\rm in}_{\omega \ell }|^2}{|\mathcal{N}^{(\lambda)\,\rm up}_{\omega \ell }|^2}|\mathcal{T}^{(\lambda)\,\text{in}}_{\omega \ell}|^2=1,\\
    |\mathcal{R}^{(\lambda)\,\text{up}}_{\omega \ell}|^2+\frac{|\mathcal{N}^{(\lambda)\,\rm up}_{\omega \ell }|^2}{|\mathcal{N}^{(\lambda)\,\rm in}_{\omega \ell }|^2}|\mathcal{T}^{(\lambda)\,\text{up}}_{\omega \ell}|^2=1.
    %\\    |\mathcal{R}^{(\lambda)\,\text{out}}_{\omega \ell m}|^2+\frac{|R_{\infty}^{(\lambda)}|^2}{|R_h^{(\lambda)}|^2}|\mathcal{T}^{(\lambda)\,\text{out}}_{\omega \ell m}|^2=1
\end{align}
The asymptotic expressions do not, by themselves, determine the origin of the mode. Instead, this is fixed by the choice of boundary conditions, through which one identifies the incident component and adopts a scattering normalization in which its amplitude is set to unity. The reflection and transmission coefficients $\mathcal{R}_{\omega \ell}^{(\lambda)}$ and $\mathcal{T}_{\omega \ell}^{(\lambda)}$ are then determined by solving the radial equations \eqref{eq: monopole_eom}, \eqref{eq: odd_parity_eom}, and \eqref{eq: even_parity_eom}. For the detailed numerical procedure, we refer the reader to Ref.\cite{Vispute:2026vek} (also see \cite{Benone:2014qaa} for earlier works on scattering). The overall normalization is fixed subsequently using the inner product, which depends on the net flux and hence on the full scattering content of the solution. Accordingly, in defining the modes, one retains only the outgoing component at the past horizon for the up-modes and only the ingoing component at null infinity for the in-modes, thereby specifying their physical origin as discussed in section \ref{sec: innerproduct and normalization}. 

One can re-express the out and down radial mode functions as linear combinations of the in- and up- radial mode functions as follows:
\begin{multline}
    u^{(\lambda)\,\text{out}}_{(i) \omega\ell}(r)=
    \mathcal{R}^{(\lambda)\,\text{in}*}_{\omega \ell} u^{(\lambda)\,\text{in}}_{(i) \omega\ell}(r) \\
    + \frac{|\mathcal{N}^{(\lambda)\,\rm in}_{\omega \ell}|^2}{|\mathcal{N}^{(\lambda)\,\rm up}_{\omega \ell }|^2}\mathcal{T}^{(\lambda)\,\text{in}*}_{\omega \ell} u^{(\lambda)\,\text{up}}_{(i) \omega\ell }(r),
\end{multline}
\begin{multline}
    u^{(\lambda)\,\text{down}}_{(i) \omega\ell}(r)=   \mathcal{R}^{(\lambda)\,\text{up}*}_{\omega \ell } u^{(\lambda)\,\text{up}}_{(i) \omega\ell}(r)\\ 
    + \frac{|\mathcal{N}^{(\lambda)\,\rm up}_{\omega \ell }|^2}{|\mathcal{N}^{(\lambda)\,\rm in}_{\omega \ell }|^2}\mathcal{T}^{(\lambda)\,\text{up}*}_{\omega \ell } u^{(\lambda)\,\text{in}}_{(i) \omega\ell }(r),
\end{multline}
where the transmission and reflection coefficient satisfy few additional properties:

\begin{align}\label{relation:ref-trans}
|\mathcal{N}^{(\lambda)\,\rm in}_{\omega \ell }|^2\mathcal{T}^{(\lambda)\,\text{in}}_{\omega \ell}&=|\mathcal{N}^{(\lambda)\,\rm up}_{\omega \ell }|^2\mathcal{T}^{(\lambda)\,\text{up}}_{\omega \ell},\\
|\mathcal{N}^{(\lambda)\,\rm in}_{\omega \ell }|^2\mathcal{R}^{(\lambda)\,\text{up}*}_{\omega \ell} \mathcal{T}^{(\lambda)\,\text{in}}_{\omega \ell} &=-|\mathcal{N}^{(\lambda)\,\rm up}_{\omega \ell}|^2\mathcal{T}^{(\lambda)\,\text{up}*}_{\omega \ell}\mathcal{R}^{(\lambda)\,\text{in}}_{\omega \ell}.
\end{align}
For deriving these equations, one needs to use the normalized radial solution. The first one can be found by equating $W(\mathcal{N}^{\rm in}_{\omega \ell }\bar{R}^{\rm in}, \mathcal{N}^{\rm up*}\bar{R}^{\rm up*})\lvert_{r=r_h}$ and $W(\mathcal{N}^{\rm in}_{\omega \ell }\bar{R}^{\rm in},\mathcal{N}^{\rm up*}\bar{R}^{\rm up*})\lvert_{r=\infty}$ while the latter can be deduced from $W(\mathcal{N}^{\rm in}\bar{R}^{\rm in}, \mathcal{N}^{\rm down*}\bar{R}^{\rm down*})\lvert_{r=r_h}$ and $W(\mathcal{N}^{\rm in}\bar{R}^{\rm in},\mathcal{N}^{\rm down*}\bar{R}^{\rm down*})\lvert_{r=\infty}$. The out-modes are constructed as specific linear combinations of the ${\rm in}$- and ${\rm up}$-modes such that no flux propagates down the event horizon. In contrast, the ${\rm down}$-modes are defined so that there is no outgoing flux at infinity. Since both the ${\rm out}$- and ${\rm down}$-modes take their simplest form on the Cauchy surface at $t \to \infty $, it is common to normalize these solutions on that hypersurface.
%%%%%%%%%%%%%%%%%%%%%%%%%%%%%%%%%%%%%%%%%%%%%%%%
\section{Inner product and the normalization of the Proca fields}\label{sec: innerproduct and normalization}
The quantization procedure relies on determining the correct symplectic structure and inner product in order to pass down the canonical commutation relations from the fields and their conjugate momenta to the creation and annihilation operators \cite{Birrell:1982ix}. To this end, a suitable definition of the inner product can be obtained by exploiting the conserved current associated with the matter action \cite{unruh1976absorption}, which allows one to construct an inner product in curved spacetime \cite{crispino2001quantization}. For the Proca field, the associated conserved current takes the following form:
\begin{equation}\label{eq: wronskian}
J^\mu(A,A') =-i\left(A_\nu F^{'\nu\mu *}- F^{\nu\mu} A_\nu^{'*}\right).
\end{equation}
Using the antisymmetry of the field-strength tensor, $F_{\mu\nu}$, and the equation of motion \eqref{eq: eom1}, it is straightforward to show that $\nabla_\mu J^\mu=0$. Therefore, using the above form of the conserved current, we define the inner product as,
\be\label{eq: inner_prod}
(A^{(\zeta)}, A^{(\zeta')})=-i\int_{\Sigma} d\Sigma_\mu J^\mu[A^{(\zeta)}, A^{(\zeta')}], 
\ee
%$$\Pi^{\mu} = \sqrt{-g}\delta^{0}_{\nu}F^{\mu\nu} = -n_{\nu}\sqrt{h}F^{\mu\nu}$$
where $d\Sigma_\mu = \epsilon\, n_\mu \sqrt{|h|}\, d^3 y$ denotes the directed surface element, with $n_\mu$ the future-directed normal, satisfying $\epsilon = n^\mu n_\mu = \pm 1$, and \{$y^i$\} with $i\equiv {1,2,3}$, representing the intrinsic coordinates on the hypersurface. Here $h$ denotes the determinant of the induced metric on the hypersurface. In this description, it is important to note that in the null limit, where the induced metric becomes degenerate and its determinant is ill-defined, the surface measure is instead constructed from the spacetime volume element and therefore involves $\sqrt{|g|}$ \cite{Poisson:2009pwt}.
%Given the static nature of the underlying spacetime, it is natural to consider $\lambda_{\mu}$ to be spacelike. However, it is sufficient to consider a Cauchy surface that lies arbitrarily close to a null surface, or a union of such surfaces.\cite{Penrose:1980yx} Beyond these requirements, we are free to choose any Cauchy surface, particularly one that simplifies the expressions obtained in \eqref{inner-prod}. %As the spacelike hypersurface $\lambda_{\mu}$ approaches the null limit, the metric determinant $\sqrt{h}=\frac{r^2\sin\theta}{\sqrt{f}}$ begins to diverge but this divergence is canceled by the vanishing unit vector $n_{\mu}(=\sqrt{f})$, so that the inner product  remains well defined. 

Importantly, given the in- and up-modes vanish at $\mathcal{H}^-$ and $\mathcal{I}^-$, respectively, it is imperative to consider $\mathcal{H}^{-}\cup\mathcal{I}^{-}$ as the Cauchy Surface, which splits the inner product for such modes as $(A^{(\zeta)}, A^{(\zeta')}) = (A^{(\zeta)}, A^{(\zeta')})_{\mathcal{H}^-} + (A^{(\zeta)}, A^{(\zeta')})_{\mathcal{I}^-}$ \cite{novikov2013physics}. Since null hypersurfaces are degenerate, no natural spacelike transverse direction exists on them. Instead, one uses a null coordinate to define the integration measure. Accordingly, to determine the normalization constant, we evaluate the inner product directly on the null hypersurfaces in their intrinsic null coordinates. For this purpose, we work in the double-null Eddington-Finkelstein coordinates, described as
\be
ds^2 = -f(r)dudv+r^2d\Omega^2\,;\qquad \sqrt{|g|}=\frac{f(r)}{2}r^2\sin\theta.
\ee
The key feature of this procedure is that the orthogonality of basis modes is independent of the explicit radial profile; it is instead fixed by the asymptotic behavior and the boundary conditions required for normalization. The precise definition of these modes are given in section \ref{sec: quantization_and_vacua}. Since the inner product is hypersurface independent, we normalize the solution on $t=-\infty$ hypersurface, where the mode is simplest. The up-mode describes an outgoing wave on the past horizon ($\mathcal{H}^{-}$), and vanishes on past null infinity ($\mathcal{I}^{-}$). To approach these boundaries suitably, we proceed by considering the following null normal vectors. In the limit $r_{*}\to -\infty$, the future directed null normal vector is given by
\be
n_{\mu}=\left(0,-1,0,0\right)
=-\delta_{\mu}^{v},\quad n^{\mu}=\frac{2}{f}\left(1,0,0,0\right)=
\frac{2}{f}\delta^{\mu}_{u}.
\ee
Using this normal vector, the inner product \eqref{eq: inner_prod} near $\mathcal{H}^-$ can be redefined as
\be\label{eq: inner_prod_Hminus}
(A^{(\zeta)}, A^{(\zeta')})_{\mathcal{H}^-}
= -i \int r^2du\,d\Omega\, J^v[A^{(\zeta)}, A^{(\zeta')}].
\ee
On the other hand, the future-directed null normal vector near $\mathcal{I}^-$ is given by
\be
n_{\mu}=\left(-1,0,0,0\right)=-\delta_{\mu}^{u},\quad n^{\mu}=\left(0,2,0,0\right)=2\delta^{\mu}_{v}.
\ee 
Therefore, near $\mathcal{I}^-$, the inner product \eqref{eq: inner_prod} takes the following form,
\be\label{eq: inner_prod_Iminus}
(A^{(\zeta)}, A^{(\zeta')})_{\mathcal{I}^-}
= -i \int r^2dv\,d\Omega\, J^u[A^{(\zeta)}, A^{(\zeta')}].
\ee
In the rest of this section, we utilize the preceding two definitions \eqref{eq: inner_prod_Hminus} and \eqref{eq: inner_prod_Iminus} to determine the normalization constants of each modes for the Proca field. To proceed, it is useful to express the components of the Proca field in the null coordinates as
\begin{align}
    &A_{u}=\frac{A_t-f(r)A_r}{2}, ~~~A_{v}=\frac{A_t+f(r)A_r}{2},\\
   % A_{u}&=\frac{u_{(1)}-u_{(2)}}{2r}Y_{\ell m} & A_{v}&=\frac{u_{(1)}+u_{(2)}}{2r}Y_{\ell m}\\
    &~~~~~~~~~~~~~~A_{\theta}=A_{\theta},
    ~~~A_{\varphi}=A_{\varphi}.
\end{align}
It is then straightforward to use the asymptotic form of the Proca field obtained in the previous section to determine the inner product among the components in the different parity sectors \eqref{eq: combined_even_basis} and \eqref{eq: combined_odd_basis}. Since the solutions behave as plane waves at the asymptotic boundaries, they can be readily expressed in $(u,v)$ coordinates using the above formula and evaluate the inner product from \eqref{eq: inner_prod_Hminus} and \eqref{eq: inner_prod_Iminus}. The detailed discussion is given in Appendix \ref{append: normalization}, we outline the main results here. For the even parity sector, the inner product between two solutions of the same polarization simplifies to,
\begin{comment}
Substituting the asymptotic form of the Proca solution, one can find from the above formulation,
\begin{align}
    A_u &= \frac{-i\omega}{q_h} (1 + i\nu r_h)\mathcal{N}^{\rm(e\pm)}_{\omega \ell }Y_{\ell m} e^{-i\omega u}\\
    A_v&=0.
\end{align}
\end{comment}
\be
\bea
&(A^{{\rm (e\pm)\,up;}\,\omega\,\ell m}, A^{{\rm (e\pm)\, up;}\,\omega'\,\ell'm'})_{\mathcal{H}^-}\nt
%&= -i \int d\lambda^u g^{\nu\alpha}\left[A_{\alpha}^{(\pm)\,\omega\,\ell m} F_{\nu u}^{(\pm)\,\omega'\,\ell'm'}^{*} - F_{\nu u}^{(\pm)\,\omega\,\ell m} A_{\alpha}^{^{(\pm)\,\omega'\,\ell'm'}*} \right]\nt %change of coordinate because $F_{uv}=0$ but $g^{uv}=\infty$
 %  &= i\int_{-\infty}^{\infty} du \int d\Omega \, r_h^2\left[ A_{u} F_{tr}^* - F_{tr} A'^{*}_u \right]\nt
 %  &\quad+i \int_{-\infty}^{\infty} du \int d\Omega \, r_h^2 g^{AB}\left[A_{A} F_{B u}^{'*} - F_{B u} A_{A}^{'*} \right]\nt
   &=\frac{4\pi \omega r_h^2}{q_h^{\pm}}(\mu^2+ \nu^2_{\pm}L)|\mathcal{N}^{\rm(e\pm)\, up}_{\omega \ell }|^2 \delta(\omega - \omega') \delta_{\ell\ell'} \delta_{mm'}.
\eea
\ee 
The norm is positive definite for $\omega>0$. In the limit $\mu\to 0$, the $\nu_{-}$ branch becomes pure gauge and has vanishing norm which agrees with massless case. Requiring $(A,A')_{\mathcal{H}^-} =\delta_{\ell\ell'}\delta_{mm'}\delta(\omega - \omega')$ fixes the normalization of these solution as

\begin{equation}
|\mathcal{N}^{\rm (e\pm)\,\rm up}_{\omega \ell }|=\frac{1}{\sqrt{\frac{4\pi |\omega| r_h^2}{q_h^{\pm}} (\mu^2+ \nu^{2}_{\pm}L)}}.
\end{equation}
\begin{comment}
The extra $r_h$ factor that appears in this normalization constant can be understood from the conserved Wronskian \eqref{conserved:wronskian}. Since the conserved quantity is the radial flux current rather than the normalization constant itself, the normalization constants are explicitly dependent on the choice of the $r=\mathrm{const.}$ hypersurface and therefore need not coincide between the horizon and spatial infinity. The radial measure factor $r^2/q_r$ entering the Wronskian is compensated by the corresponding hypersurface-dependent normalization coefficients, ensuring that the total flux remains conserved under radial evolution. 
\end{comment}
Following the similar procedure, the normalization constant for odd parity up-modes is found to be
\be
\mathcal{N}^{\rm ({\rm o})\, up}_{\omega \ell }=\sqrt{\frac{L}{4\pi|\omega|}}.
\ee 
Notably, since no flux reaches $\mathcal{I}^{+}$, the bound-state solutions for $\omega<\mu$ admit a normalization analogous to the $\omega>\mu$ case, as the procedure is independent of this separation of regimes. 

The solution orthogonal to up- modes on the same Cauchy surface are in- modes. They originate from the past null infinity as purely ingoing waves and vanish on $\mathcal{H}^{-}$. This simplifies the inner product, and we can directly evaluate the inner product over $\mathcal{I}^{-}$ to normalize the solution. For the in-modes of even parity, the inner product can be expressed as follows:

\begin{comment}
This fixes the mode function as:
\begin{align}
u_{(1)}(\omega, r) &\xrightarrow{r \to \infty} -\frac{k}{\nu_{\pm}} \bar{R}^{(e\pm)}_{\rm in}, \\
u_{(2)}(\omega, r) &\xrightarrow{r \to \infty} -\frac{\omega}{\nu_{\pm}}\bar{R}^{(e\pm)}_{\rm in}.
\end{align}
\end{comment}
\be
\bea    
&(A^{{\rm (e\pm)\, in;}\,\omega\,\ell m}, A^{{\rm (e\pm)\, in;}\,\omega'\,\ell'm'})_{\mathcal{I}^-}\\ &= 4\pi k\left(\frac{\mu^2}{\nu^2_{\pm}} +L\right)|\mathcal{N}^{\rm (e\pm)\, in}_{\omega \ell }|^2 \delta(\omega - \omega') \delta_{\ell\ell'} \delta_{mm'}.\label{in--mode-innerprod}
\eea
\ee
%For $\omega\gtrless0$, we have $p,K\gtrless0$ and thus the sign of inner product depends on the $\omega$. 
Notably, for $\omega<\mu$, the solution is not normalizable as it diverges at the null infinity. Therefore, we restrict $\omega>\mu$ and require $(A,A')_{\mathcal{I}^-} =\delta_{\ell\ell'}\delta_{mm'}\delta(\omega - \omega')$ for in- modes, which then fixes the normalization as:

\begin{equation}
|\mathcal{N}_{\omega \ell }^{(e\pm)\,\rm in}| = \frac{|\nu_{\pm}|}{\sqrt{4\pi k(\mu^2+\nu^2_{\pm}L)}}.
\end{equation}
Here, taking the limit $\mu\to0$ of \eqref{eq: inner_prod} is subtle. Unlike the case of up-modes where the final expression remained well-defined in the massless limit, the inner product here becomes indeterminate because the asymptotic limit $r\to\infty$ does not commute with the $\nu_-\to0$ limit. Since the $\nu_-$ branch becomes pure gauge in the massless limit, consistency requires taking the $\nu_-\to0$ limit first in order to ensure that the pure-gauge mode indeed carries vanishing norm. 

Following the same procedure, one can derive the normalization constant for in- modes in the odd parity sector as,
\be
\mathcal{N}^{\rm ({\rm o})\,\rm in}_{\omega \ell }=\sqrt{\frac{L}{4\pi k}}.
\ee
Whereas, even parity polarizations $e+$ and $e-$ are found to be orthogonal to each other under \eqref{eq: inner_prod}.
\be
(A^{(e+)\, \rm in}_{\omega\,\ell m}, A^{(e-)\,\rm in}_{\omega\,\ell m})=0,~~~~
%=\frac{\left(\mu^2+\nu_{+}\nu_{-}L\right)}{\sqrt{(\mu^2+\nu^2_{+}L)(\mu^2+\nu^2_{-}L)}}
(A^{(e+)\, \rm up}_{\omega\,\ell m}, A_{\omega\,\ell m}^{(e-)\,\rm up})=0.\label{cross}
\ee
Therefore, the full orthonormality relation is given by
\be
(A^{(\lambda)\,n;\,\omega\,\ell m}, A^{(\lambda')\,n';\,\omega'\,\ell'm'}) =\delta^{\lambda\lambda'}\delta_{nn'}\delta_{\ell\ell'}\delta_{mm'}\delta(\omega-\omega').
\ee
\section{Canonical Quantization and the structure of Vacua}\label{sec: quantization_and_vacua}
%%%%%%%%%%%%%%%%%%%%%%%%%%%%%%%%%%%%%%%
Having discussed the construction of the orthonormal classical modes via the inner product in the previous section, the commutator algebra can now be derived directly from the fields and their associated conjugate momenta. As we shall see, however, a naive implementation of this procedure does not yield the correct commutation relations. To see where the problem arises, let us begin with the Lagrangian density of the Proca field corresponding to the action in \eqref{eq: proca_action},
\be
\mathcal{L}=-\frac14 F_{\mu\nu}F^{\mu\nu}-\frac12 \mu^2 A_\mu A^\mu.
\ee
The conjugate momenta density is found to be
\be\label{eq: conj_momentum}
\Pi^\mu=\pdv{\sqrt{-g}\mathcal{L}}{(\partial_0A_{\mu})}=\sqrt{-g}F^{\mu0}.
\ee
An important feature appears immediately, that for the temporal component, conjugate momentum vanishes, i.e., $\hat\Pi^0=0$. As a result of this primary constraint \cite{dirac2013lectures}, one can not directly impose 
\be
[\hat A^0(x),\hat\Pi^0(y)]_{x^0=y^0}=ig^{00}\delta(\bf{x}-\bf{y}').
\ee 
A similar situation arises for the electromagnetic field as well. Generically, in the framework of a gauge field, one typically either adds a gauge fixing term in the action \cite{crispino2001quantization}, which prevents $\Pi^0$ from vanishing or fixes the constraints at the level of the equation of motion (such as the implementation of Coulomb gauge). For the latter case, with the inconsistency in the above commutator relation, promoting Poisson brackets straightforwardly to commutators does not generate the consistent algebra.

To proceed in this situation, we follow the Dirac bracket formalism \cite{Dirac:1950pj, Bergmann:1949zz, Henneaux:1994lbw, Date:2010xr, Rothe:2010dzf, Wipf:1993xg, Brown:2022gha, Vathsan:1995mp, Bergmann:1954tc}(for a simple introduction to this topic see e.g., \cite{das2020lectures, Date:2010xr}) for quantizing the massive vector field in a BH background as opposed to reduced phase space formalism \cite{Chingangbam:1999jkm, Pavel:1997pi}, where we integrate out $A_0$ and then quantize the whole system as often done in the context of Early Universe Cosmology (see \cite{Graham:2015rva, Kolb:2020fwh}). However, since the algebraic structure becomes intuitive once the temporal components are fixed, one may express it in terms of the canonical momentum using the equation of motion of the Proca field \eqref{eq: eom1} as
\be
\hat{A}^0=\frac{\partial_i \hat{\Pi}^i}{\mu^2\sqrt{-g}}\label{A0constraint}.
\ee
Now, we present here the commutators of the dynamical variables as follows: 
\begin{align}\label{eq: commutator_set1}
&\bigl[\hat{\Pi}^i(x),\hat{\Pi}^j(y)\bigr]_{x^0=y^0}=0,\\&\bigl[\hat{A}^i(x),\hat{A}^j(y)\bigr]_{x^0=y^0}=0, \label{eq: commutator_set2}\\
&\bigl[\hat{A}^i(x),\hat{\Pi}^j(y)\bigr]_{x^0=y^0}=ig^{ij}\,\delta^{3}(\mathbf{x}-\mathbf{y}).\label{eq: commutator_set3}
\end{align}
and defer the detailed derivation to Appendix \ref{append: diracbracket}.
%For other combinations,
The remaining Dirac brackets are automatically satisfied and do not require separate treatment, as can be realized by utilizing \eqref{A0constraint} and the above commutators %\eqref{eq: commutator_set1}-\eqref{eq: commutator_set3},

\begin{align}
\bigl[\hat A^0(x),\hat{\Pi}^j(y)\bigr]_{x^0=y^0}&=\frac{1}{\mu^2\sqrt{-g}}\bigl[\partial_i\hat{\Pi}^i(x),\hat{\Pi}^j(y)\bigr]_{x^0=y^0}=0,\\
\bigl[\hat A^0(x),\hat{A}^j(y)\bigr]_{x^0=y^0}
%&=\frac{1}{\mu^2\sqrt{-g}}\bigl[\partial_i\hat{\Pi}^i(x),\hat{A}^j(y)\bigr]_{x^0=y^0}\notag\\
%&=\frac{1}{\mu^2\sqrt{-g}}\partial^i\underbrace{\bigl[\hat{\Pi}_i(x),\hat{A}^j(y)\bigr]_{x^0=y^0}}_{-i\delta_{i}^j\delta^{3}(\mathbf{x}-\mathbf{y})}\notag\\
&=-\frac{i}{\mu^2\sqrt{-g}}\,\partial^j\delta^{3}(\mathbf{x}-\mathbf{y}).
\end{align}
An important consequence of the above algebraic structure is that although the field operator may be decomposed as $\hat A_\mu=(\hat A_0,\hat A_i)$,
the temporal component is not an independent dynamical degree of freedom. As shown in \eqref{A0constraint}, the operator $\hat A_0$ is completely determined by the spatial sector. Hence, it inherits the creation and annihilation operators from $\hat{\Pi}^i$ and thereby $\hat{A_i}$. %Hence, the quantized Proca field does not contain an independent unphysical timelike polarization degree of freedom. 

In constructing the field expansion, an important property of the inner product \eqref{eq: inner_prod} is that a mode is orthogonal to its complex conjugate, i.e., $(A_{\,\omega \ell m}, A_{\,\omega \ell m}^{*})=0$.
%  &= -i \int_{\Sigma_t}\bigg[A^{\nu}_{\omega \ell m}F^{\omega \ell m}_{\nu\mu}- F^{\omega \ell m}_{\nu\mu}A^{\nu}_{\omega \ell m}\bigg] d\Sigma^{\mu}\nt=0,
Hence, one can decompose $\hat{A}$ in the orthonormal basis of $A_{\mu}^{\zeta}$ and $A_{\mu}^{\zeta*}$
\be
\bea    
\hat{A}_{\mu}&= \int_{0}^{\mu}\,d\omega\sum_{\lambda\,\ell m}\hat a_{\omega\,\ell m}^{(\lambda)} A_{\mu\,\omega \ell m}^{(\lambda)\,\rm up}(x)+a_{\omega\,\ell m}^{(\lambda)\dagger} A_{\mu\,\omega \ell m}^{(\lambda)\,\rm up*}(x)\\
&~+\sum_{n}\int_{\mu}^{\infty}\,d\omega\sum_{\lambda\,\ell m}\hat a_{\omega\,\ell m}^{(\lambda)\,n} A_{\mu\,\omega \ell m}^{(\lambda)\,n}(x)+a_{\omega\,\ell m}^{(\lambda)\,n\dagger} A_{\mu\,\omega \ell m}^{(\lambda)\,n*}(x)\label{mode-decom}.
\eea
\ee
In the first line of the above expansion, representing the bound state regime, $\omega<\mu$, only up-modes appear as discussed previously. Whereas, in the regime of scattering states, $\omega>\mu$, both in- and up-modes must be taken into account. This is included through the summation over $n$, which, as mentioned before, provides the directionality of the modes. With this setup, the mode expansion for the conjugate momentum can readily be expressed, using the definition \eqref{eq: conj_momentum}, as
\be
\bea    
\hat{\Pi}^{\mu}&= \int_{0}^{\mu}\,d\omega\sum_{\lambda\,\ell m}\hat a_{\omega\,\ell m}^{(\lambda)} \Pi_{\omega \ell m}^{(\lambda)\mu}(x)+a_{\omega\,\ell m}^{(\lambda)\dagger} \Pi_{\omega \ell m}^{(\lambda)\mu*}(x)\\
&~+\sum_{n}\int_{\mu}^{\infty}\,d\omega\sum_{\lambda\,\ell m}\hat a_{\omega\,\ell m}^{(\lambda)} \Pi_{\,\omega \ell m}^{(\lambda)\mu}(x)+a_{\omega\,\ell m}^{(\lambda)\dagger} \Pi_{\,\omega \ell m}^{(\lambda)\mu*}(x)\label{pi-mode-decom},
\eea
\ee
where the directionality follows the field, $A^\mu$. Next, to pass down the commutator structure of the fields and their conjugate momentum to the creation and annihilation operators, let us consider the following combination:
\begin{equation}
    \int d^3x \left[\Pi^{(\lambda)\,\mu}_{\omega \ell m}(t,x)^* \hat A_{\mu}(t,x) - A^{(\lambda)\mu}_{\omega \ell m}(t,x)^* \hat\Pi_{\mu}(t,x)\right].\label{eq: consider}
\end{equation}
Substituting \eqref{mode-decom} and \eqref{pi-mode-decom} in above and using the normalization condition
\begin{multline}
    \int d^3x \left[\Pi^{(\lambda)*}_{\mu\,\omega \ell m} A^{(\lambda')\, \mu}_{\omega' \ell' m'}
    - A^{(\lambda)\,\mu*}_{\omega \ell m} \Pi^{(\lambda' )}_{\mu \omega' \ell' m'}\right]\\
=i\delta^{\lambda \lambda'}\delta_{\ell\ell'}\delta_{mm'}\delta(\omega-\omega'),
\end{multline}
we obtain
%The LHS becomes:
\begin{multline}\label{eq: annihilation_op}
    \int d^3x \left[\Pi^{(\lambda)\,\mu}_{\omega \ell m}(t,x)^* \hat A_{\mu}(t,x) - A^{(\lambda)\mu}_{\omega \ell m}(t,x)^* \hat\Pi_{\mu}(t,x)\right]\\
    =i\delta^{\lambda \lambda'}\delta_{\ell\ell'}\delta_{mm'}\int d\omega'\,\delta(\omega-\omega')\hat a_{\omega'\ell'm'}^{(\lambda')}.
\end{multline}
In order to write the RHS of \eqref{eq: consider}, 
%in standard form
one needs to use \eqref{eq: inner_prod} with the conjugate momentum expressed in the following manner
\begin{equation}
    \Pi_{\mu} %= \sqrt{-g}g^{0\nu}\hat F_{\mu\nu} 
    = \epsilon\,n^{\nu}\sqrt{h} F_{\mu\nu}, \label{Pi1}
\end{equation}
where $n^\mu$ denotes the future-directed timelike unit normal to the hypersurface $\Sigma_t$, $h$ is the determinant of the induced spatial metric on $\Sigma_t$, and $\epsilon=n^\mu n_\mu=-1$.
Then, one directly obtains from \eqref{eq: annihilation_op} the following formula for the annihilation operator,
    \begin{align}
        \hat a^{(\lambda)}_{\omega\ell m} &= -i \int d^3x \left[\hat A_{\mu} \Pi^{(\lambda)\,\mu}_{\omega \ell m}\,^* -\hat \Pi^{\mu}A_{\omega \ell m}^{(\lambda)\,\mu}\,^*\right]\nt
    &=(\hat A(t,x),A_{\omega \ell m}^{(\lambda)}).\label{a1}
    \end{align}

\begin{comment}
Where $n\in\{\rm in, up\}$ the modes are decomposed as:
\begin{align}
A_{\mu \,\omega \ell m}^{(e+)\, \rm n}&=%\frac{i}{\omega}\left(\frac{f(r)}{r}u_{(2)}^{\ell m}+\partial_{r*}u_{(2)}^{\ell m}\right)Z^{(1)\ell m}_{\mu}+u_{(2)}^{\ell m}Z^{(2)\ell m}_{\mu}\\
\frac{1}{r}\sum^3_{i=1}\sum_{\ell m}c_{(i)}u^{(e+)\, \rm n\,\ell }_{(i)}(t,r) Z^{(i)\ell m}_{\mu}(r, \Omega)\\
A_{\mu \,\omega \ell m}^{(e-)\, \rm n}&=\frac{1}{r}\sum^3_{i=1}\sum_{\ell m}c_{(i)}u^{(e-)\, \rm n\,\ell }_{(i)}(t,r) Z^{(i)\ell m}_{\mu}(r, \Omega)\\
%\frac{f(r)}{ir\omega}u_{(3)}^{\ell m}Z^{(1)\ell m}_{\mu}+c_{(4)}u_{(3)}^{\ell m}Z^{(3)\ell m}_{\mu}\\
A_{\mu \,\omega \ell m}^{({\rm o})\, \rm n}&=c_{(4)}\frac{u_{(4)}^{({\rm o})\, \rm n\,\ell }(t,r)}{r}Z^{(4)\ell m}_{\mu}
\end{align}
\end{comment}
%To consistently perform the canonical quantization of the system, one requires that the commutator algebra be correctly induced on the creation and annihilation operators. This is achieved by combining the equal-time commutation relations of the fields with the definition of the creation and annihilation operators in terms of the Proca inner product as follows (see Appendix \ref{append: inner_prod}),
Therefore, the commutator of the creation and annihilation operator boils down to 
\begin{align}
    [\hat a^{(\lambda)}_{\omega \ell m}, \hat a^{(\lambda' )\dagger}_{\omega' \ell' m'}] &=[(\hat A(t,x),A_{\omega \ell m}^{(\lambda)}),(\hat A(t,x),A_{\omega' \ell' m'}^{(\lambda')})^{\dagger}]\nt
    &= (A^{(\lambda' )}_{\omega' \ell' m'}, A^{(\lambda)}_{\omega \ell m}).
\end{align}
The analysis up to this point closely parallels the flat spacetime treatment \cite{greiner2013field}. The essential difference in the BH backgrounds is the absence of a globally defined timelike Killing vector, which renders the notion of particles and vacuum observer dependent \cite{Birrell:1982ix}. The non-uniqueness begins with the choice of positive-frequency modes. Defining positive frequency modes with respect to the Schwarzschild time $t$ is natural for asymptotic observers, whereas regularity across the horizon requires analyticity in the Kruskal coordinates $U$ and $V$. These choices lead to inequivalent vacuum states with distinct physical interpretations \cite{Das:2019aii}. In Schwarzschild spacetime, it gives rise to the three standard vacua: the Boulware, Unruh, and Hartle--Hawking states. 
%Another motivation for considering different vacua in curved spacetime is the requirement that $\langle T_{\mu\nu}\rangle$ stays finite at $r=r_h$ \cite{Balbinot:2023vcm}. 
Below, we provide the definition of the three vacuum states usually defined in the BH background and discuss the associated mode functions relevant for the present analysis. 

%A particularly important feature of BH spacetimes is that the Kubo--Martin--Schwinger (KMS) periodicity condition for two-point correlation function, $G(t+i\beta)=\eta\, G(t)$ with $\beta=\kappa/2\pi$, and $\eta=\pm1$ for bosons and fermions, emerges naturally from the analytic structure of the field modes in Kruskal coordinates.
%%%%%%%%%%%%%%%%%%%%%%%%%%%%%%%%%%%%%%%%
\subsection{Boulware Vacuum}
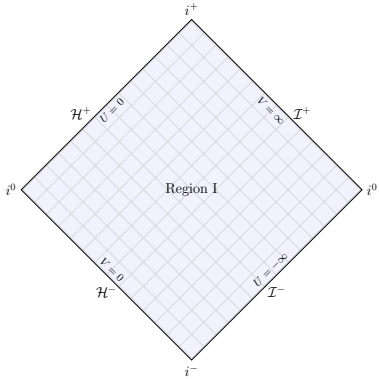
\begin{figure}[h]
    \centering
    \resizebox{0.6\linewidth}{!}{
    \colorlet{fillI}{blue!5}
        \begin{tikzpicture}
        \fill[fillI] (0,-5) -- (5,0) -- (0,5) -- (-5,0) -- cycle;% I
        \node[fill=fillI, inner sep=1pt] at (1.1,0) {};
            % 1. The Grid (Clipped to the diamond)
            \begin{scope}
                \clip (0,-5) -- (5,0) -- (0,5) -- (-5,0) -- cycle;
                \begin{scope}[rotate=45]
                    \draw[gray!30, step=0.5] (-5,-5) grid (5,5);
                \end{scope}
            \end{scope}

            % 2. The Main Boundaries
            \draw[thick]
            (0,-5) -- node[midway, above, sloped] {$U=-\infty$} (5,0)
                -- node[midway, below, sloped] {$V=\infty$} (0,5)
                -- node[midway, below, sloped] {$U=0$} (-5,0)
                -- node[midway, above, sloped] {$V=0$} cycle;

            % 3. Labels
            \large
            \node[right] at (5,0) {$i^0$};
            \node[above] at (0,5) {$i^+$};
            \node[below] at (0,-5) {$i^-$};
            \node[left] at (-5,0) {$i^0$};
            
    		\node at (0.0, 0.) {Region I};
            \node at (2.5,-3) {$\mathcal{I}^-$};
            \node at (3.25,2.25) {$\mathcal{I}^+$};
            \node at (-2.5,-3) {$\mathcal{H}^-$};
            \node at (-3.25,2.25) {$\mathcal{H}^+$};
        \end{tikzpicture}
    }
    \caption{Penrose diagram from Region I.}
    \label{fig:minkowski_penrose}
\end{figure}
For asymptotic observers, the positive frequency modes are defined with respect to the timelike Killing vector of the Schwarzschild geometry. These modes are characterized by
\begin{align}
\mathcal{L}_{\partial_t} f_{\omega}=-i\omega f_{\omega},
\qquad \omega>0,
\end{align}
throughout the exterior region $r>2GM$ (see Fig.~\eqref{fig:minkowski_penrose}). Quantization with respect to these modes defines the Boulware vacuum \cite{Boulware:1974dm}. By construction, this state contains no incoming radiation from $\mathcal{I}^{-}$ and no outgoing flux at $\mathcal{I}^{+}$, and therefore corresponds to the vacuum perceived by static observers at infinity.

The field operator on $\mathcal{I}^-$ may then be expanded in a complete basis of in-modes
\[\left\{A_{\mu\,\omega \ell m}^{(\lambda)\,\rm in},A_{\mu\,\omega \ell m}^{(\lambda)\,\rm in*}\right\},
\]
which satisfy the orthonormality relations
\begin{align}
\left(
A_{\omega \ell m}^{(\lambda)\,\rm in},
A_{\omega' \ell' m'}^{(\lambda')\,\rm in}
\right)
&=
\delta^{\lambda\lambda'}
\delta_{\ell\ell'}
\delta_{mm'}
\delta(\omega-\omega'),
\\
\left(
A_{\omega \ell m}^{(\lambda)\,\rm in*},
A_{\omega' \ell' m'}^{(\lambda')\,\rm in*}
\right)
&=
-
\delta^{\lambda\lambda'}
\delta_{\ell\ell'}
\delta_{mm'}
\delta(\omega-\omega').
\end{align}
Similarly, the set
\[
\left\{
A_{\mu\,\omega \ell m}^{(\lambda)\,\rm up},
A_{\mu\,\omega \ell m}^{(\lambda)\,\rm up*}
\right\}
\]
forms a complete orthonormal basis on the past horizon $\mathcal{H}^{-}$, satisfying analogous normalization conditions. The field operator in Region~I may therefore be expanded in terms of both the in- and up- sectors as
\begin{multline}
\hat{A}_{\mu}=\int_{\mu}^{\infty} d\omega\sum_{\lambda \ell m}\left[\hat a_{\omega \ell m}^{(\lambda)}A_{\mu\,\omega \ell m}^{(\lambda)\,\rm in}+\hat a_{\omega \ell m}^{(\lambda)\dagger}A_{\mu\,\omega \ell m}^{(\lambda)\,\rm in*}\right]\\
+\int_{0}^{\infty} d\omega\sum_{\lambda \ell m}\left[\hat b_{\omega \ell m}^{(\lambda)}A_{\mu\,\omega \ell m}^{(\lambda)\,\rm up}+\hat b_{\omega \ell m}^{(\lambda)\dagger}A_{\mu\,\omega \ell m}^{(\lambda)\,\rm up*}\right].
\end{multline}
The Boulware vacuum $\ket{0_B}$ is then defined through

\begin{align}
\hat a_{\omega \ell m}^{(\lambda)} \ket{0_B}=0,
\qquad
\hat b_{\omega \ell m}^{(\lambda)} \ket{0_B}=0,
\end{align}
for all $\omega,\lambda,\ell,m$, and the creation and annihilation operators obey:

\begin{align}
     [a^{(\lambda)}_{\omega\, \ell m},a_{\omega'\, \ell'm'}^{(\lambda' )\dagger}]&=\delta(\omega-\omega')\delta^{\lambda\lambda'}\delta_{\ell\ell'}\delta_{mm'},\\
    [b_{\omega\, \ell m}^{(\lambda)},b_{\omega'\, \ell'm'}^{(\lambda' )\dagger}]&=\delta(\omega-\omega')\delta^{\lambda\lambda'}\delta_{\ell\ell'}\delta_{mm'}.
\end{align}
Here, the rest of the commutators vanish. An identical construction may be carried out on $\mathcal{I}^{+}\cup\mathcal{H}^{+}$ by replacing the in- and up- bases with
\[
\left\{
A_{\mu\,\omega \ell m}^{(\lambda)\,\rm out},
A_{\mu\,\omega \ell m}^{(\lambda)\,\rm out*}
\right\},
\qquad
\left\{
A_{\mu\,\omega \ell m}^{(\lambda)\,\rm down},
A_{\mu\,\omega \ell m}^{(\lambda)\,\rm down*}
\right\},
\]
defined on $\mathcal{I}^{+}$ and $\mathcal{H}^{+}$ respectively. The corresponding vacuum state defines the future Boulware vacuum.

By construction, the Schwarzschild notion of positive frequency becomes ill-defined at the event horizon. As a result, while the mode functions remain regular at spatial infinity, the renormalized stress tensor diverges near the future horizon $\mathcal{H}^{+}$. Consequently, the Boulware vacuum fails to define a regular quantum state on the BH horizon.
%%%%%%%%%%%%%%%%%%%%%%%%%%%%%%%%%%%%%%%%%%%%%%%%%%%%%%%%%%%%%%%%%%%%%%%%%%
\subsection{Unruh Vacuum}
\begin{figure}
    \centering
    \includegraphics[width=\linewidth]{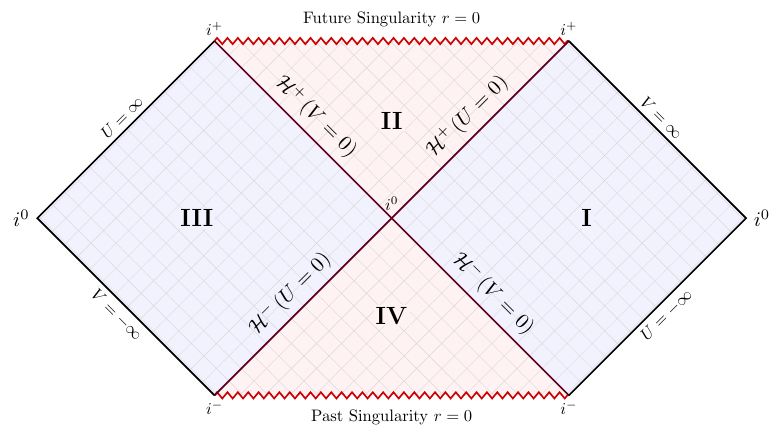}
    \caption{Penrose diagram of maximally extended Schwarzschild spacetime.}
    \label{fig:placeholder}
\end{figure}

Unlike the Boulware vacuum, the Unruh vacuum is not defined by positive
frequency with respect to the Schwarzschild Killing time everywhere.\cite{Unruh:1976db} The Schwarzschild coordinates become singular at the horizon, and therefore regularity of the quantum state requires positive frequency with respect to the affine Kruskal coordinates on the corresponding null horizon.

For the past Unruh vacuum, the in-modes remain positive frequency modes on past null infinity \(\mathcal I^{-}\),

\be
A_{\mu}^{\rm in}\sim e^{-i\omega v},\qquad \omega>0,
\ee
where $v=t+r_\ast$. These modes are naturally associated with the asymptotic time translation generated by \(\partial_t\). The nontrivial sector is the up-modes, since these originate from the past horizon
\(\mathcal H^{-}\). Such modes must instead be positive frequency with
respect to the affine Kruskal coordinate \(U\). These coordinates are defined as \cite{Kruskal:1959vx},
\be
U=-\frac{1}{\kappa}e^{-\kappa u},~~~V=\frac{1}{\kappa}e^{\kappa v},
\ee
with $u=t-r_\ast$. Then the outgoing near horizon modes in region I behave as
\begin{align}
\bar R_{\omega\ell}^{(I)\lambda\,\rm up}e^{-i\omega t}&\sim e^{-i\omega u}=e^{i\frac{\omega}{\kappa}\ln(-\kappa U)}\Theta(-U).
\label{unruh:I}
\end{align}
These up-modes are only defined in region I and region IV corresponding to \(U<0\). The mode function on the second independent branch can be given as,
\begin{align}
\bar R_{\omega\ell}^{(III)\lambda\,\rm up}e^{-i\omega t}&=e^{i\frac{\omega}{\kappa}\ln(\kappa U)}\Theta(U),
\label{unruh:III}
\end{align}
which is defined only in region II and region III. The second branch mode function is related to \eqref{unruh:I} via the \(\mathbb Z_2\) isometry of the maximally extended Schwarzschild geometry (see Fig.~\ref{fig:placeholder}),
\begin{align}
(U, V)\rightarrow(-U,-V),\label{Z2}
\end{align}
which maps region I to region III. To determine which linear combinations correspond to positive frequency with respect to the affine Kruskal coordinate \(U\), one uses the analyticity criterion. A function is positive frequency with respect to \(U\) if it admits the Fourier decomposition
\begin{align}
f(U)=\int_0^\infty dp\, c(p)e^{-ipU}.
\end{align}
Equivalently, its Fourier transform has no support on the opposite frequency-branch,
\begin{align}
\int_{-\infty}^{\infty} dU\, e^{-ipU}f(U)=0,\qquad p>0.
\end{align}
This condition is equivalent to analyticity of $f(U)$ and boundedness in the lower half complex \(U\)-plane.

The required analytic continuation follows from the standard lemma that
for positive real \(p\) and arbitrary real \(q\) \cite{Novikov:1989sz}
\be
\bea
&\int_{-\infty}^{\infty} dX\, e^{-ipX}\left[e^{-iq\ln X}\Theta(X)+e^{-\pi q}e^{-iq\ln(-X)}\Theta(-X)\right]\\
&~=0,\label{analyticlemma}
\eea
\ee
provided the branch of $\log$ is chosen such that the integrand is analytic on lower half plane.

Using Eq.~\eqref{analyticlemma}, one finds that the unique positive frequency combination with respect to the affine coordinate \(U\) is
\begin{align}
\bar R_{\omega\ell}^{(\lambda)\,\rm up+}&=\frac{e^{\frac{\pi\omega}{2\kappa}}
\bar R_{\omega\ell}^{(I)\lambda\,\rm up}+e^{-\frac{\pi\omega}{2\kappa}}\bar R_{\omega\ell}^{(III)\lambda\,\rm up}}{\sqrt{2\sinh(\pi\omega/\kappa)}},\label{analytic:upplus}
\end{align}
while the orthogonal combination,
\begin{align}
\bar R_{\omega\ell}^{(\lambda)\,\rm up-}&=\frac{e^{-\frac{\pi\omega}{2\kappa}}\bar R_{\omega\ell}^{(I)\lambda\,\rm up}+e^{\frac{\pi\omega}{2\kappa}}\bar R_{\omega\ell}^{(III)\lambda\,\rm up}}{\sqrt{2\sinh(\pi\omega/\kappa)}},\label{analytic:upminus}
\end{align}
is analytic in the upper half-plane and therefore corresponds to
negative Kruskal frequency.

Using these globally analytic modes, the field operator may be expanded
as
\be
\bea
\hat A_\mu &=\sum_{\lambda\ell m}\int_{\mu}^{\infty} d\omega\left[\hat a_{\omega\ell m}^{(\lambda)}A_{\mu\,\omega\ell m}^{(\lambda)\,\rm in}+\hat a_{\omega\ell m}^{(\lambda)\dagger}A_{\mu\,\omega\ell m}^{(\lambda)\,\rm in*}\right]\\
&~~~~+\sum_{\lambda\ell m}\int_0^\infty d\omega\left[\hat b_{\omega\ell m}^{(\lambda)}A_{\mu\,\omega\ell m}^{(\lambda)\,\rm up+}+\hat b_{\omega\ell m}^{(\lambda)\dagger}
A_{\mu\,\omega\ell m}^{(\lambda)\,\rm up-}\right].
\eea
\ee
The past Unruh vacuum is then defined by
\be
\hat a_{\omega\ell m}^{(\lambda)}
\ket{0_U}
=\hat b_{\omega\ell m}^{(\lambda)}
\ket{0_U}=0.
\ee
Thus the state contains no incoming particles from \(\mathcal I^{-}\) but contains an outgoing flux of Hawking radiation at future null infinity \(\mathcal{I}^+\) while regular on the future horizon \(\mathcal H^{+}\), it diverges on the past horizon $\mathcal{H}^-$.
The future Unruh vacuum is constructed analogously by imposing analyticity of the down-modes across the future horizon
\(\mathcal H^{+}\). In that case, the affine coordinate \(V\) replaces \(U\) throughout the above analysis.
%%%%%%%%%%%%%%%%%%%%%%%%%%%%%%%%%
\subsection{Hartle--Hawking State}
The Hartle--Hawking state is defined as the unique regular thermal state on the maximally extended Schwarzschild spacetime~\cite{Hartle:1976tp}. In contrast to the Unruh vacuum, where analyticity is imposed only on the outgoing sector associated with the affine coordinate \(U\), the Hartle--Hawking construction requires analyticity in both Kruskal coordinates \((U, V)\). As a result of this analyticity, both the outgoing and ingoing sectors satisfy the Kubo--Martin--Schwinger (KMS) condition \cite{Fulling:1987otn},
\be
G^{+}(t-i\beta_H)=G^{-}(t),
\qquad
\beta_H=\frac{2\pi}{\kappa},
\ee
corresponding to a thermal state at the Hawking temperature $T_H=\kappa/(2\pi)$. Physically, this describes a black hole in thermal equilibrium with an incoming bath of Hawking quanta.

The analytic continuation of the outgoing up- modes proceeds identically to the Unruh construction discussed previously. One must additionally impose the same analyticity condition on the ingoing in- modes across the future horizon \(\mathcal H^{+}\). The near-horizon behavior of the in-modes in region I is therefore

\begin{align}
\bar R_{\omega\ell}^{(I)\lambda\,\rm in}e^{-i\omega t}
=
e^{-i\omega v}
=
e^{-i\frac{\omega}{\kappa}\ln(\kappa V)}\Theta(V).
\end{align}
Under the \(\mathbb Z_2\) isometry
\begin{align}
(U,V)\rightarrow(-U,-V),
\end{align}
the corresponding branch in region III becomes
\begin{align}
\bar R_{\omega\ell}^{(III)\lambda\,\rm in}e^{-i\omega t}
=
e^{-i\frac{\omega}{\kappa}\ln(-\kappa V)}\Theta(-V).
\end{align}
\begin{comment}
The positive-frequency Kruskal modes are obtained by analytically continuing through the lower complex $V$-plane,
\begin{align}
v\rightarrow v-\frac{i\pi}{\kappa},
\end{align}
which ensures convergence for $\omega>0$. Choosing the opposite branch instead produces negative-frequency solutions and leads to ultraviolet divergence in the Fourier transform with respect to $V$. 
\end{comment}
Following the same procedure as before, the globally analytic combinations are therefore
\begin{align}
\bar R_{\omega\ell}^{\rm in+}&=\frac{e^{\frac{\pi\omega}{2\kappa}}\bar R_{\omega\ell}^{(I)\lambda\,\rm in}+e^{-\frac{\pi\omega}{2\kappa}}\bar R_{\omega\ell}^{(III)\lambda\,\rm in}}{\sqrt{2|\sinh(\pi\omega/\kappa)|}},\label{analytic: in+}\\
\bar R_{\omega\ell}^{\rm in-}&=\frac{e^{-\frac{\pi\omega}{2\kappa}}\bar R_{\omega\ell}^{(I)\lambda\,\rm in}+e^{\frac{\pi\omega}{2\kappa}}\bar R_{\omega\ell}^{(III)\lambda\,\rm in}}{\sqrt{2|\sinh(\pi\omega/\kappa)|}}.\label{analytic: in}
\end{align}
Together with the analytically extended up-modes defined previously, the field operator admits the expansion
\begin{widetext}
\begin{multline}
\hat A_\mu=\sum_{\lambda\ell m}\int_{|\omega|>\mu} d\omega\left[\hat a_{\omega\ell m}^{(\lambda)}A_{\mu\,\omega\ell m}^{(\lambda)\,\rm in+}+\hat a_{\omega\ell m}^{(\lambda)\dagger}A_{\mu\,\omega\ell m}^{(\lambda)\,\rm in-}\right]+\sum_{\lambda\ell m}\int_{-\infty}^{\infty}d\omega\left[\hat b_{\omega\ell m}^{(\lambda)}A_{\mu\,\omega\ell m}^{(\lambda)\,\rm up+}+\hat b_{\omega\ell m}^{(\lambda)\dagger}A_{\mu\,\omega\ell m}^{(\lambda)\,\rm up-}\right].
\end{multline}
In region I, this reduces to
\begin{multline}
    \hat A_\mu=\sum_{\lambda\ell m}\int_{|\omega|>\mu}d\omega\,\frac{A_{\mu\,\omega\ell m}^{(\lambda)\,\rm in}}{\sqrt{|2\sinh(\pi\omega/\kappa)|}}\left[e^{\frac{\pi\omega}{2\kappa}}\hat a_{\omega\ell m}^{(\lambda)}+e^{-\frac{\pi\omega}{2\kappa}}\hat a_{\omega\ell m}^{(\lambda)\dagger}\right]
    +\sum_{\lambda\ell m}\int_{-\infty}^{\infty}d\omega\,\frac{A_{\mu\,\omega\ell m}^{(\lambda)\,\rm up}}{\sqrt{|2\sinh(\pi\omega/\kappa)|}}\left[e^{\frac{\pi\omega}{2\kappa}}\hat b_{\omega\ell m}^{(\lambda)}+e^{-\frac{\pi\omega}{2\kappa}}\hat b_{\omega\ell m}^{(\lambda)\dagger}\right].
\end{multline}
\end{widetext}
The Hartle--Hawking vacuum is then defined by
\be
\hat a_{\omega\ell m}^{(\lambda)}\ket{0_{\rm HH}}=\hat b_{\omega\ell m}^{(\lambda)}\ket{0_{\rm HH}}=0.
\ee
Contrary to Unruh vacuum, Hartle Hawking state describes a thermal state at $\mathcal{I}^+$ and $\mathcal{I}^-$, while remaining regular on $\mathcal{H}^+$ and $\mathcal{H}^-$. We now move on to show how Hawking radiation appears from the vacuum structure defined above. 

%%%%%%%%%%%%%%%%%%%%%%%%%%%%%%%%%%%%%%%%%%%%%%%%%%%%%%%%%%%%%%%%%%%%%%%%%%%
\section{Calculation of Hawking Flux}\label{sec: hawkingflux}
The asymptotic observer at future null infinity $\mathcal{I}^+$ defines a basis of positive-frequency physical out- modes $A^{(\lambda),\mathrm{out}}_{\omega \ell m}$. These modes are normalized on $\mathcal{I}^+$ according to
\be
(A^{(\lambda)\,\text{out}}_{\omega \ell m}, \, A^{(\lambda)\,\text{out}}_{\omega' \ell m})_{\mathcal{I}^{+}} = \delta(\omega - \omega').
\ee
The analysis is performed in the Heisenberg picture, where the quantum state is fixed and the dynamics is encoded in the field operators. In curved spacetime, however, the decomposition of the field operator into positive- and negative-frequency sectors is observer dependent. As a result, the annihilation operators associated with the asymptotic past basis on $\mathcal{I}^- \cup \mathcal{H}^-$ need not be the same on $\mathcal{I}^+ \cup \mathcal{H}^+$. 

To characterize the Hawking flux, we choose the vacuum $\ket{0}$, which is defined to be empty with respect to positive-frequency in-modes on $\mathcal{I}^{-}$. Although the state itself does not evolve in the Heisenberg picture, the notion of vacuum defined by future asymptotic observers differs from that defined in the asymptotic past. Consequently, the expectation value of the future number operator in the state $|0\rangle$ becomes nonvanishing, corresponding to a thermal population of outgoing particles at $\mathcal{I}^{+}$.

The out- modes may therefore be expanded in terms of the complete basis on $\Sigma_{\mathrm{past}} = \mathcal{I}^- \cup \mathcal{H}^-$, consisting of the in- and up-modes,
\begin{multline}\label{eq: out_decomp}
    A^{(\lambda)\,\text{out}}_{\omega \ell m} = \int_{\mu}^{\infty} d\omega'' \bigg[ \alpha_{\omega \omega''}^{\text{out,in}\,(\lambda)} A^{(\lambda)\,\text{in}}_{\omega'' \ell m} + \beta_{\omega \omega''}^{\text{out,in}\,(\lambda)} A^{(\lambda)\,\text{in}*}_{\omega'' \ell m} \bigg]\\
    +  \int_{0}^{\infty} d\omega''\bigg[\alpha_{\omega \omega''}^{\text{out,up}\,(\lambda)} A^{(\lambda)\,\text{up}}_{\omega'' \ell m} + \beta_{\omega \omega''}^{\text{out,up}\,(\lambda)} A^{(\lambda)\,\text{up}*}_{\omega'' \ell m} \bigg],
\end{multline}
For the subsequent analysis, we will keep this limit of the $\omega-$integral implicit and suppress the polarization label $\lambda$ to simplify the notation wherever necessary. Nevertheless, the conservation of the inner product leads to
\be
(A^{(\lambda)\,\text{out}}_{\omega \ell m}, \, A^{(\lambda)\,\text{out}}_{\omega' \ell m})_{\mathcal{I}^{-}}\, +\, (A^{(\lambda)\,\text{out}}_{\omega \ell m}, \, A^{(\lambda)\,\text{out}}_{\omega' \ell m})_{\mathcal{H}^{-}} = \delta(\omega - \omega').
\ee
Upon substitution the integral \eqref{eq: out_decomp} becomes
\be\label{eq: inner_prod_in_out}
\bea
&\int d\omega'' \bigg( \alpha_{\omega \omega''}^{\text{out,in}} \alpha_{\omega' \omega''}^{\text{out,in}*} - \beta_{\omega \omega''}^{\text{out,in}} \beta_{\omega' \omega''}^{\text{out,in}*} + \alpha_{\omega \omega''}^{\text{out,up}} \alpha_{\omega' \omega''}^{\text{out,up}*}\\
&~~~~~~~~~~~~~~~~~- \beta_{\omega \omega''}^{\text{out,up}} \beta_{\omega' \omega''}^{\text{out,up}*} \bigg) = \delta(\omega - \omega'),
\eea
\ee
with the Bogoliubov coefficients: 
\begin{equation}
    \begin{aligned}
        \alpha_{\omega \omega''}^{n, n'\,(\lambda)} &= (A^{(\lambda)\, n'}_{\omega \ell m}, \, A^{(\lambda)\, n}_{\omega'' \ell m})_{\mathcal{I}^{-}\cup\mathcal{H}^{-}},\\
\beta_{\omega \omega''}^{n, n'\,(\lambda)} &= -(A^{(\lambda)\, n'}_{\omega \ell m}, \, A^{(\lambda)\,n^*}_{\omega'' \ell m})_{\mathcal{I}^{-}\cup\mathcal{H}^{-}}.
    \end{aligned}\label{alpha-beta}
\end{equation}
Now we impose the classical boundary conditions. Given the static nature of the underlying spacetime, if we trace the physical out-mode backward in time from $\mathcal{I}^+$ to $\mathcal{I}^-$, it is simply the fraction of the wave that classically reflected off the potential. 
\begin{figure*}[t]
%    \centering
\includegraphics[width=\textwidth]{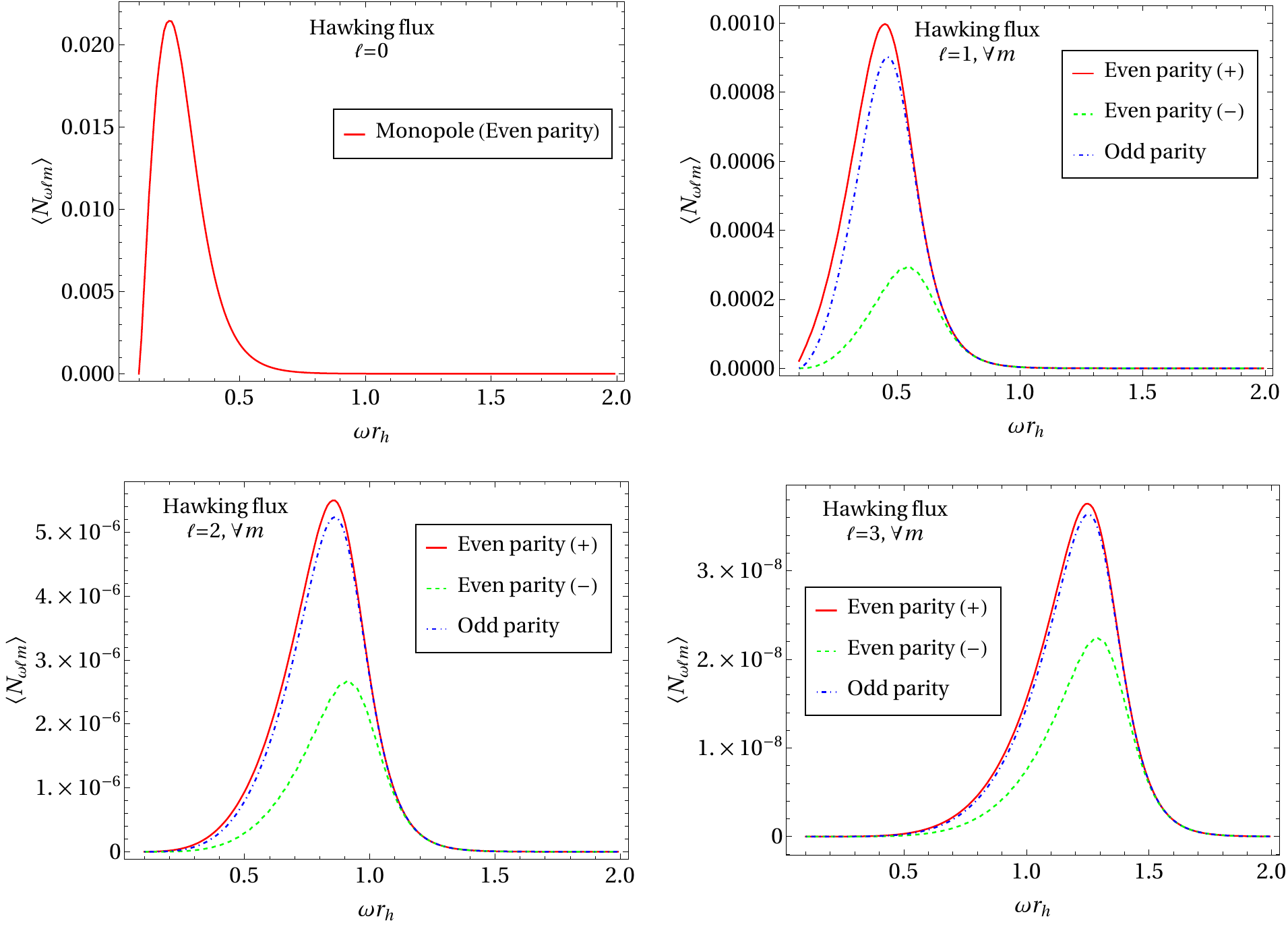}
\caption{The Hawking spectrum has been plotted for different polarizations. Comparisons have been made between different $\ell$ modes at $\mathcal{I}^+$.}
\label{fig:basiscomparison}
\end{figure*}
Because classical scattering off a static background does not mix positive and negative frequencies, the reflected wave is purely positive-frequency. Using \eqref{alpha-beta}, the Bogoliubov coefficients mapping to the in--mode are given by
\be
\alpha_{\omega \omega''}^{\text{out,in}\,(\lambda)}=\mathcal{R}_{\omega\ell}^{(\lambda)\,\rm in\,*} \delta(\omega - \omega''),~~~~\beta_{\omega \omega''}^{\text{out,in}\,(\lambda)}= 0.
\ee
It further simplifies the inner product \eqref{eq: inner_prod_in_out}, which can now be expressed as, 
\be
\bea
&\int_{\mu}^{\infty} d\omega''|\mathcal R_{\omega\ell}^{(\lambda)\,\rm up}|^2 \delta(\omega - \omega'')\delta(\omega' - \omega'')\\ 
&+\int_{0}^{\infty} d\omega''\bigg(  \alpha_{\omega \omega''}^{\text{out,up}} \alpha_{\omega' \omega''}^{\text{out,up}*} - \beta_{\omega \omega''}^{\text{out,up}} \beta_{\omega' \omega''}^{\text{out,up}*} \bigg) = \delta(\omega - \omega').
\eea
\ee
\begin{comment}
By using equation \eqref{ref-trans-normalization}, we can express it as
\begin{multline}
\int_0^\infty d\omega'' \left( \alpha_{\omega \omega''}^{\text{out,up}} \alpha_{\omega' \omega''}^{\text{out,up}*} - \beta_{\omega \omega''}^{\text{out,up}} \beta_{\omega' \omega''}^{\text{out,up}*} \right) \\
= \frac{|\mathcal{N}^{(\lambda)\,\rm up}_{\omega \ell }|^2}{|\mathcal{N}^{(\lambda)\,\rm in}_{\omega \ell }|^2}|\mathcal T_{\omega\ell}^{(\lambda)\,\rm up}|^2 \delta(\omega - \omega').
\end{multline}
\end{comment}
This can be simplified using
\begin{equation}
|\alpha_{\omega \omega'}^{\text{out,up}\,(\lambda)}| = e^{\frac{\pi\omega}{\kappa}} |\beta_{\omega \omega'}^{\text{out,up}\,(\lambda)}|.
\end{equation}
One can derive{\footnote{The structure of the analysis is similar to the scalar case with identical behaviour of the modes near the boundary. For this, we refer the reader to Chapter 4 of Ref.\cite{Parker:2009uva} for scalar field.} this by using the analytically continued near-horizon form of the up-modes and utilising \eqref{alpha-beta}. Similarly, one can find
\begin{multline}
    \hat{a}_{\omega \ell m}^{\text{out}\,(\lambda)} = \int d\omega''\bigg( \alpha_{\omega \omega''}^{\text{out,in}\,(\lambda)*} \hat{a}_{\omega''\ell m}^{\text{in}\,(\lambda)} - \beta_{\omega \omega''}^{\text{out,in}\,(\lambda)*} \hat{a}_{\omega''\ell m}^{\text{in}\,(\lambda)\dagger} \\
    + \alpha_{\omega\omega''}^{\text{out,up}\,(\lambda)*} \hat{a}_{\omega''\ell m}^{\text{up}\,(\lambda)} - \beta_{\omega \omega''}^{\text{out,up}\,(\lambda)*} \hat{a}_{\omega''\ell m}^{\text{up}\,(\lambda)\dagger} \bigg).
\end{multline}
Next we calculate the particle density on $\mathcal{I}^{+}$
\begin{align}
    \bra{0}N^{\rm out\,(\lambda)}_{\omega\,\ell m}\ket{0}=\int_0^\infty d\omega' |\beta_{\omega \omega'}^{\text{out,up}\,(\lambda)}|^2
%    &= \sum_{\lambda}|(M^{-\frac{1}{2}})\,_{\lambda}\,^{\lambda}|^2\frac{|\mathcal{N}^{(\lambda)\,\rm up}_{\omega \ell }|^2}{|\mathcal{N}^{(\lambda)\,\rm in}_{\omega \ell }|^2}\frac{|\mathcal T_\omega^{(\lambda)\,\rm up}|^2}{e^{\frac{2\pi\omega}{\kappa}}-1}\nt
= \frac{\left(1-|\mathcal R_{\omega\ell}^{(\lambda)\,\rm up}|^2\right)}{e^{\frac{2\pi\omega}{\kappa}}-1}.
\end{align}
The spectrum is thermal with greybody factor $(1-|\mathcal R_{\omega\ell}^{(\lambda)\,\rm up}|^2)$. Notably, for $\omega < \mu$, the transmission coefficient vanishes, and therefore there is no flux at $\mathcal{I}^{+}$ in this regime. To find where this missing flux goes, one needs to consider what happens to number operator near the horizon. To find the particle density crossing the future event horizon $\mathcal{H}^+$, we can similarly expand the physical down-modes $A^{(\lambda)\,\text{down}}_{\omega \ell m}$ in the past basis.
Then following the procedure as above particle density crossing $\mathcal{H}^{+}$ in the vacuum state can be given as,
\be
\bra{0}N^{\rm down\,(\lambda)}_{\omega\,\ell m}\ket{0}=\int_0^\infty d\omega' |\beta_{\omega \omega'}^{\text{down,up}\,(\lambda)}|^2=\frac{|\mathcal{R}_{\omega\ell}^{(\lambda)\,\rm up}|^2}{e^{\frac{2\pi\omega}{\kappa}}-1}.
\ee
This result shows that the flux that was missing at $\mathcal{I}^+$ ($\omega < \mu$), is localized near the horizon.

The key distinction between the massless \cite{Page:1976df} and massive vector cases is the presence of an additional physical degree of freedom corresponding to the
longitudinal polarization of the Proca field. Numerically, we observe that the
${\rm e(+)}$, even-parity sector, and the odd-parity sector remain nearly
degenerate even in the massive case, as can be seen from Fig.~\ref{fig:basiscomparison}. Whereas the additional ${\rm e(-)}$ sector
exhibits a comparatively weaker Hawking flux. This suggests that the extra
longitudinal degree of freedom predominantly manifests through this component. The comparatively smaller flux associated with this sector is also consistent with the general suppression of massive particle emission in Hawking radiation. As
the angular momentum number \(\ell\) increases, the centrifugal barrier suppresses the overall flux for all sectors. Nevertheless, the ${\rm e-}$ mode appears to become comparatively less suppressed relative to the other polarizations at larger \(\ell\), indicating a nontrivial polarization dependence in the Hawking spectrum of Proca field.
%%%%%%%%%%%%%%%%%%%%%%%%%%%%%%%%%%%%%%
\section{Vacuum Expectation Values of two point functions and the Proca Condensate}\label{sec: vevs}
%%%%%%%%%%%%%%%%%%%%%%%%%%%%%%%%%%
The expectation values of the quadratic operator are fully determined by the following two-point function \cite{crispino2001quantization},
\begin{equation}
G^{\Psi}_{\mu\nu'}(x,x')=\bra{\Psi}\hat A_\mu(x)\hat A_{\nu'}(x')\ket{\Psi},
\end{equation}
for $\ket{\Psi}\in\{\ket{0_B},\ket{0_U},\ket{0_{HH}}\}$. A particularly useful local observable is the Proca condensate, defined as 
\be
\langle \hat A_\mu \hat A^\mu \rangle_\Psi=\lim_{x'\to x}g^{\mu\nu'}G^{\Psi}_{\mu\nu'}(x,x').
\ee
Being a scalar in nature, the condensate provides a coordinate-independent description of vacuum polarization effects in the BH background. Furthermore, it can be used directly to compare the behaviour of different quantum states near the future horizon and in the asymptotic region. Following the discussion on the past vacuum states from, the two-point functions corresponding to the Boulware, Unruh, and Hartle--Hawking vacua are given by
\begin{widetext}
\begin{align}
G^{B}_{\mu\nu'}(x,x')%\bra{0_B}\hat A_\mu(x)\hat A_{\nu'}(x')\ket{0_B}
=\sum_{\lambda,\ell,m}\bigg[&\int_{\mu}^{\infty} d\omega\;A^{(\zeta)\,\mathrm{in}}_{\mu}(x)\,A^{(\zeta)\,\mathrm{in}*}_{\nu'}(x')+\int_{0}^{\infty} d\omega\;A^{(\zeta)\,\mathrm{up}}_{\mu}(x)\,A^{(\zeta)\,\mathrm{up}*}_{\nu'}(x')\bigg],\label{GB}\\
G^{U}_{\mu\nu'}(x,x')%\bra{0_U}\hat{A}_\mu(x)\hat{A}_{\nu'}(x')\ket{0_U}
=\sum_{\lambda,\ell,m}\bigg[&\int_{\mu}^{\infty} d\omega\;A^{(\zeta)\,\mathrm{in}}_{\mu}(x)\,A^{(\zeta)\,\mathrm{in}*}_{\nu'}(x')+\int_{0}^{\infty} d\omega\;\coth\left(\frac{\pi\omega}{\kappa}\right)A^{(\zeta)\,\mathrm{up}}_{\mu}(x)\,A^{(\zeta)\,\mathrm{up}*}_{\nu'}(x')\bigg],\label{GU}\\
G^{HH}_{\mu\nu'}(x,x')%\bra{0_{HH}}\hat{A}_{\mu}(x)\hat{A}_{\nu'}(x')\ket{0_{HH}}
=\sum_{\lambda,\ell,m}\bigg[&\int_{\mu}^{\infty} d\omega\;\coth\!\left(\frac{\pi\omega}{\kappa}\right)A^{(\zeta)\,\mathrm{in}}_{\mu}(x)\,A^{(\zeta)\,\mathrm{in}*}_{\nu'}(x')+\int_{0}^{\infty} d\omega\;\coth\!\left(\frac{\pi\omega}{\kappa}\right)A^{(\zeta)\,\mathrm{up}}_{\mu}(x)\,A^{(\zeta)\,\mathrm{up}*}_{\nu'}(x')\bigg].\label{GH}
\end{align}
\end{widetext}
Notably, the in- and up-modes are displayed along with $\zeta$. Nevertheless, the above two-point functions contain the usual ultraviolet divergences associated with the coincidence limit of local quantum field operators \cite{Christensen:1976vb}. Similar to a scalar field, one may construct finite observables by subtracting the Boulware contribution as follows \cite{Balakumar:2022yvx},
\be
\Delta G_{\mu\nu}^{\Psi-B}=\bra{\Psi}\hat A_\mu \hat A_\nu \ket{\Psi}-\bra{0_B}\hat A_\mu \hat A_\nu \ket{0_B}.
\ee
This subtraction removes the universal local divergent structure of the correlator. The resulting expression is ultraviolet finite, since the thermal factor suppresses the integrand exponentially for $\omega\gg\kappa$. Consequently, the remaining nontrivial behaviour arises from the infrared region, which we will discuss for each quantum state and one must therefore analyze the $\omega\to0$ limit separately for each quantum state.
%%%%%%%%%%%%%%%%%%%%%%%%%%%%%%%%%%%%%%%
\subsection{Future Horizon}
We now study the behaviour of the Proca condensate near the future event horizon $\mathcal{H}^{+}$. Since the Unruh vacuum is regular on the future horizon, the corresponding condensate provides a natural probe of the vacuum fluctuation experienced by infalling observers. The relative condensate with respect to the Boulware vacuum is given by (for detailed discussion see Appendix \ref{append: twopointasymptotic})
\begin{widetext}
    \begin{align}
        \langle \hat{A}_\mu \hat{A}^\mu \rangle_{U-B}
        &= \sum_{\lambda, \ell} \frac{2\ell+1}{4\pi r^2} \int_{0}^{\infty} d\omega \left[ \frac{2}{e^{\frac{\pi\omega}{\kappa}}-1}\right] \left( \frac{-|u_1^{(\lambda)\,\rm up}|^2+|u_2^{(\lambda)\,\rm up}|^2 }{f}+ \frac{|u_3^{(\lambda)\,\rm up}|^2 + |u_4^{(\lambda)\,\rm up}|^2}{L}  \right)\label{AU-B},\\
        \langle \hat{A}_\mu \hat{A}^\mu \rangle_{HH-B} &= \sum_{n\in\{\rm in, up\}}\sum_{\lambda, \ell} \frac{2\ell+1}{4\pi r^2} \int_{\omega_{\rm min}}^{\infty} d\omega \left[ \frac{2}{e^{\frac{\pi\omega}{\kappa}}-1}\right] \left( \frac{-|u_1^{(\lambda)\,\rm n}|^2+|u_2^{(\lambda)\,\rm n}|^2 }{f}+ \frac{|u_3^{(\lambda)\,\rm n}|^2 + |u_4^{(\lambda)\,\rm n}|^2}{L}  \right).\label{AHH-B}
    \end{align}
\end{widetext}
To integrate the above equation, one needs to use the frequency-dependent numerical solution and further perform the integration numerically. Given the complexity, we estimate and evaluate the above expression in the near-horizon limit where $|u_{(1)}^{(e\pm)}|=|u_{(2)}^{(e\pm)}|$. Following this procedure, we obtain,
\be
\bea
\langle \hat{A}_\mu \hat{A}^\mu \rangle_{U-B}&\sim\sum_{\lambda, \ell} \frac{2\ell+1}{4\pi r^2L} \int_{0}^{\infty} d\omega \left[ \frac{2}{e^{\frac{\pi\omega}{\kappa}}-1}\right] \\
&~~~~~~~~~\times\left(|u_3^{(\lambda)\,\rm up}|^2 + |u_4^{(\lambda)\,\rm up}|^2\right),\label{AUrh}\\
\eea
\ee
and 
\be
\bea
\langle \hat{A}_\mu \hat{A}^\mu \rangle_{HH-B}&\sim\sum_{n\in\{\rm in, up\}}\sum_{\lambda, \ell} \frac{2\ell+1}{4\pi r^2L} \int_{\omega_{\rm min}}^{\infty} d\omega \left[ \frac{2}{e^{\frac{\pi\omega}{\kappa}}-1}\right]\\
&~~~~~~\times\left(|u_3^{(\lambda)\,\rm n}|^2 + |u_4^{(\lambda)\,\rm n}|^2 \right). \label{AHrh}
\eea
\ee
An interesting consequence of the above expressions is that the monopole mode, which resides entirely in the longitudinal polarization branch $(\nu_-)$, does not contribute to the condensate at $r=r_h$. In contrast, this mode contributes nontrivially in the asymptotic region, as we discuss in the next section. Moreover, since the radial solutions become purely oscillatory near the horizon, the phase factors cancel out. Consequently, the residual radial dependence in \eqref{AUrh} and \eqref{AHrh} arises solely from the overall factor of $1/r^2$. Nevertheless, the integrand of \eqref{AHrh} diverges in the infrared limit $\omega\to 0$. This divergence originates from using the Boulware vacuum as the reference state and reflects its singular behavior at the horizon. It is therefore more physical to consider the condensate relative to the Unruh vacuum, which remains regular on $\mathcal{H}^{+}$. The corresponding expression can be written as
\begin{widetext}
    \begin{align}
        \langle \hat{A}_\mu \hat{A}^\mu \rangle_{HH-U} &= \sum_{\lambda, \ell} \frac{2\ell+1}{4\pi r^2} \int_{\mu}^{\infty} d\omega \left[ \frac{2}{e^{\frac{\pi\omega}{\kappa}}-1}\right] \left( \frac{-|u_1^{(\lambda)\,\rm in}|^2+|u_2^{(\lambda)\,\rm in}|^2 }{f}+ \frac{|u_3^{(\lambda)\,\rm in}|^2 + |u_4^{(\lambda)\,\rm in}|^2}{L}  \right).\label{AHH-U}
    \end{align}
\end{widetext}
Near the horizon $(r\to r_h)$, it becomes
\begin{align}
    \langle \hat{A}_\mu \hat{A}^\mu \rangle_{HH-U}
        &\sim\sum_{\lambda, \ell} \frac{2\ell+1}{4\pi r^2L} \int_{\mu}^{\infty} d\omega \left[ \frac{2}{e^{\frac{\pi\omega}{\kappa}}-1}\right]\nt
        &\qquad\times\left(|u_3^{(\lambda)\,\rm in}|^2 + |u_4^{(\lambda)\,\rm in}|^2 \right).\label{AHUrh}
\end{align}
\begin{figure}[!t]
\includegraphics[scale=0.55]{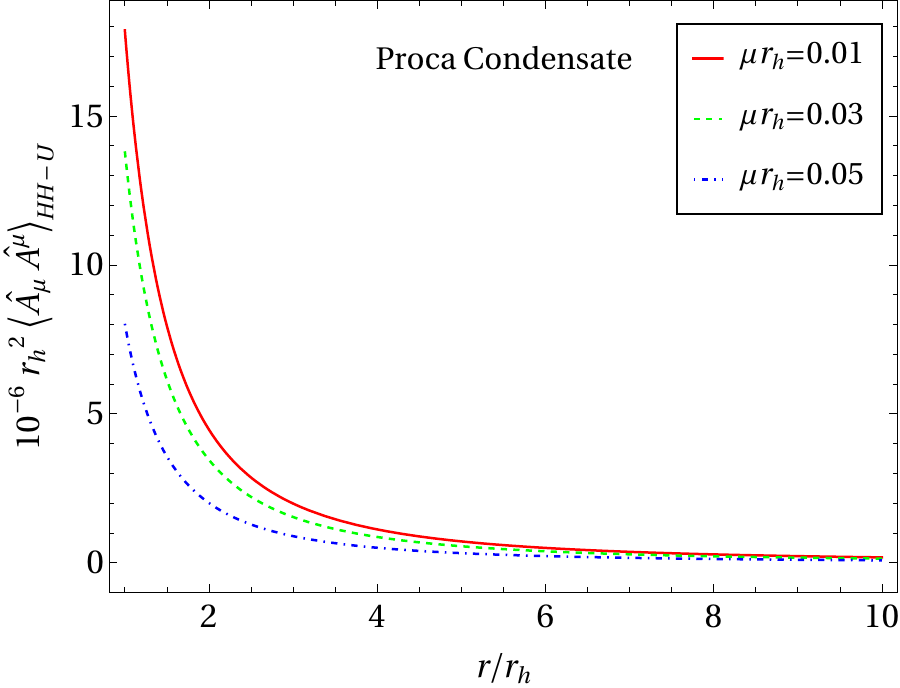}
\caption{The Proca condensate has been plotted with respect to the radial distance for three different field masses. This condensate is measured by a stationary observer near the future horizon.}
\label{fig: proca_cond}
\end{figure}
In evaluating the integral, one should notice that the expression is exponentially suppressed for high frequencies. On the other hand, concerning the summation over the $\ell$ modes, transmission for in-modes becomes very small as we increase $\ell$, specifically in the low frequencies, and also dominated by the $\ell=1$ mode (barring the monopole mode). Therefore, we evaluate the above expression by numerically performing the integration for the first few $\ell$ modes. In Fig.~\ref{fig: proca_cond}, the behaviour of the condensate has been illustrated for three different closely spaced values of Proca mass, $\mu$. The mass-dependency of the condensate, manifested in the figure, can be understood from the expression of the effective potentials (see \Cref{sec: classical_soln}), by realizing the increase of the barrier height with $\mu$ for all the parity modes, which naturally leads to the in-modes being transmitted in lesser amounts.
%%%%%%%%%%%%%%%%%%%%%%%%%%%
\subsection{Future Null Infinity}
We have seen that the monopole mode, in which the additional longitudinal degree of freedom has its dominant contribution, is absent from the near-horizon condensate. This naturally motivates studying the condensate in the asymptotic region, where the longitudinal sector can contribute nontrivially, as reflected in the Hawking flux at $\mathcal{I}^{+}$. Moreover, physical observables associated with Hawking radiation are ultimately measured by asymptotic observers at future null infinity rather than by local probes near $\mathcal{H}^{+}$. It is therefore important to examine the asymptotic behavior of the condensate and compare how the different quantum states encode the outgoing thermal flux.

Taking the asymptotic limit $(r\to\infty)$ of \eqref{AHH-B},\eqref{AU-B} and \eqref{AHH-U}, we obtain
\begin{widetext}
\begin{align}
\langle \hat{A}_\mu \hat{A}^\mu \rangle_{U-B}&\sim \sum_{\lambda, \ell} \frac{2\ell+1}{4\pi r^2} \int_{0}^{\infty} d\omega \left[ \frac{2}{e^{\frac{\pi\omega}{\kappa}}-1}\right] \left(-|u_1^{(\lambda)\,\rm up}|^2+|u_2^{(\lambda)\,\rm up}|^2 + \frac{|u_3^{(\lambda)\,\rm up}|^2 + |u_4^{(\lambda)\,\rm up}|^2}{L}  \right),\label{AUinf}\\
\langle \hat{A}_\mu \hat{A}^\mu \rangle_{HH-B}&\sim \sum_{n\in\{\rm in, up\}}\sum_{\lambda, \ell} \frac{2\ell+1}{4\pi r^2} \int_{\omega_{\rm min}}^{\infty} d\omega \left[ \frac{2}{e^{\frac{\pi\omega}{\kappa}}-1}\right] \left(-|u_1^{(\lambda)\,\rm n}|^2+|u_2^{(\lambda)\,\rm n}|^2 + \frac{|u_3^{(\lambda)\,\rm n}|^2 + |u_4^{(\lambda)\,\rm n}|^2}{L}\right),\label{AHrinf}\\
\langle \hat{A}_\mu \hat{A}^\mu \rangle_{HH-U}&\sim \sum_{\lambda, \ell} \frac{2\ell+1}{4\pi r^2} \int_{\mu}^{\infty} d\omega \left[ \frac{2}{e^{\frac{\pi\omega}{\kappa}}-1}\right] \left(-|u_1^{(\lambda)\,\rm in}|^2+|u_2^{(\lambda)\,\rm in}|^2 + \frac{|u_3^{(\lambda)\,\rm in}|^2 + |u_4^{(\lambda)\,\rm in}|^2}{L}  \right).\label{AHUrinf}
    \end{align}
\end{widetext}
Notably, the asymptotic structure of the up-modes at $\mathcal{I}^+$ depends on the transmission coefficients $\mathcal{T}^{(\lambda)}_{\omega\ell}$ (see \Cref{sec: asymptoticform}), rendering the $\omega\to 0$ limit well defined. In contrast, for the in-modes, the condition $\omega>\mu$ structurally removes any potential infrared divergence. This is consistent with the fact that none of these vacuum states is singular at asymptotic null infinity.

The two-point functions and the associated condensates 
\(
\langle \hat{A}_\mu \hat{A}^\mu \rangle_{\Delta}
\)
provide a local description of quantum fluctuations in the different vacuum states of the Proca field on a Schwarzschild background. In particular, they reveal how the near-horizon geometry suppresses the longitudinal monopole contribution, while the asymptotic region remains sensitive to the additional Proca degree of freedom. The comparison between the Boulware, Unruh, and Hartle--Hawking states further makes explicit how the infrared structure of the condensate is tied to the regularity of the underlying vacuum at the horizon. These observables, therefore, provide a useful probe of vacuum polarization and quantum-state dependence for massive vector fields in curved spacetime.
%%%%%%%%%%%%%%%%%%%%%%%%%%%%%%%%%%%%%%%%%
\section{Conclusion and outlook}\label{sec: conclusion}
In the present work, we have presented a framework for the canonical
quantization of a massive vector field in a Schwarzschild BH background, and explored the significance of the mass term in the ultimate quantum mechanical observables. The procedure begins with the evaluation of classical mode functions, where we have introduced a combination of VSH and FKKS basis to handle the field components based on earlier classical analysis of the Proca field \cite{Fernandes:2021qvr, Rosa:2011my}. Moving forward to the quantization steps, a central difficulty of the Proca theory, absent in the scalar
case \cite{Balakumar:2022yvx, Egorov:2022hgg}, is the presence of second-class constraints associated with the vanishing of the timelike component of conjugate momenta \cite{dirac2013lectures, Brown:2022gha}. To overcome this step, we
implemented the Dirac--Bergmann algorithm \cite{Brown:2022gha, Date:2010xr}, explicitly identified the symplectic structure of the constrained system, and constructed the
 %promoted the resulting Dirac brackets to
commutator algebra. 
%We further demonstrated in detail how the equal-time commutation relations of the fields are inherited by the corresponding creation and annihilation operators through the Proca inner product.

Using the orthonormalized basis, we define the Boulware, Unruh, and Hartle--Hawking vacua and, as a primary observable, derived Hawking flux and two-point correlation functions. The Hawking spectrum exhibits several characteristic features unique to the Proca field due to the presence of an additional physical longitudinal polarization. We observed that the exact degeneracy between the \(e+\) and \(o\) modes, expected in the massless limit, is mildly lifted for a massive Proca field. Nevertheless, the two sectors remain remarkably close numerically even for finite mass. In the massless limit, the $e-$ polarization becomes pure gauge. On the other hand, in the massive case, it is a physical degree of freedom. Importantly, the monopole mode being contained entirely in this new degree of freedom, the analogue of its dynamical behaviour is completely absent in the massless case. The total Hawking flux is dominated by the monopole mode, with higher modes being suppressed due to the centrifugal barrier. In the latter case, we also find that polarization of $e-$ is relatively suppressed compared to $e+$ and $o$ polarizations. Nevertheless, the relative suppression of the $e-$ sector becomes weaker at larger $\ell$, indicating a nontrivial polarization dependence in the Proca greybody factors \cite{Bunjusuwan:2025enh}.

The Proca mass $\mu$ also introduces a physically important threshold in the
spectrum. For frequencies satisfying $\omega>\mu$, the field exhibits
standard Hawking radiation together with the additional longitudinal emission
channel absent in Maxwell theory. By contrast, modes with $\omega<\mu$ are
unable to propagate to null infinity due to the effective gravitational
potential barrier. Instead, these low-frequency excitations remain localized
near the BH and behave as trapped Proca fluctuations, potentially
relevant for the accumulation of ultralight dark photon configurations around BHs \cite{Rosa:2011my}.

We further analyzed differences of expectation values between distinct quantum
states, which provide finite observables without the need for explicit
renormalization (see \cite{Balakumar:2022yvx} for the scalar field). In particular, near the horizon, or asymptotically, we find that
\begin{equation}
    \langle \hat{A}_{\mu}\hat{A}^{\mu}\rangle_{HH-B},\qquad\langle \hat{A}_{\mu}\hat{A}^{\mu}\rangle_{U-B},
\end{equation}
exhibit a leading-order behavior proportional to \(1/r^{2}\) with complicated radial dependence in the intermediate region. The strong
near-horizon enhancement reflects the growth of quantum fluctuations in the
thermal states relative to the Boulware vacuum. The vacuum states and two-point functions constructed here provide the necessary starting point for future computations of the renormalized stress-energy tensor \(\langle T_{\mu\nu}\rangle\) \cite{Frolov:1989jh}. Such analyses are essential for understanding the semiclassical backreaction of massive vector fields on BH geometries.

Based on the present work, several directions can be explored in future. A natural extension is the quantization of the Proca field in rotating BH backgrounds, where one must carefully derive the appropriate symplectic structure and analyze the role of superradiant modes in the construction of physical vacua \cite{Unruh:1974bw}. Another important concern arises in the consistent definition and renormalization of quantum states inside the BH horizon (\(r<r_h\)) \cite{Lanir:2017oia, Lanir:2018rap}. 
\begin{comment}
On the other hand, the framework developed here may serve as a starting point for studying interacting massive vector theories in curved spacetime, including the effects of self-interactions \cite{}, non-minimal couplings, and the massless limit in the presence of residual gauge ambiguities.
In particular, understanding the behaviour of the longitudinal polarization and the structure of correlation functions across the horizon may provide further insight into the quantum properties of massive vector fields in strong gravitational backgrounds. The scattering and absorption of Proca fields by black holes also deserve further study, especially the polarization dependence of the corresponding greybody factors and quasinormal spectra. Since the monopole sector is dominated by the longitudinal mode, it would be interesting to investigate whether this feature leads to observable signatures in black hole evaporation or modifies the late-time behavior of massive vector perturbations. 
\end{comment}

{\bf Acknowledgements :}
We thank Kaustubh Mukund Vispute for various discussions on the classical Proca field. RK gratefully acknowledges discussions with Debaprasad Maity, Chandramouli Chowdhury, Ashmita Das and Mahajani Rohan Sachindra on quantization in curved spacetime over the years. RK would also like to thank Xian-Hui Ge and his research group for the continuous support and encouragement at Shanghai University. CP would like to acknowledge Prathamesh Changde for his assistance with several TikZ diagrams during the preparation of the manuscript.
%%%%%%%%%%%%%%%%%%%%%%%%%%%%%%%%%%%%%%%%%
\appendix
\renewcommand{\thesection}{\Alph{section}}
\renewcommand{\theequation}{\thesection.\arabic{equation}}
\section{Derivation of the normalization factors}\label{append: normalization}
Let us first consider the even parity sector $(\rm e\pm)$. In the near-horizon limit, the equation governing the even parity modes can be expressed as
\begin{align}
\left[\pr^2_{r_*}+\omega^2\right]\bar{R}^{\rm (e\pm)}_{\rm up}(\omega,r)=0.
\end{align}
The up- modes are governed by the radial solution of the form: $\bar{R}_{\rm up}(\omega, r) = e^{i\omega r_*}$, which leads to $f \partial_r \bar{R}_{\rm up} = i\omega \bar{R}_{\rm up}$. On $v=-\infty$, $r_{*}\to-\infty$ and $r\to r_{h}\quad\rm (const)$.
The asymptotic limit as $r \to r_h$ yields
\be
\bea
&u_{(1)}^{(e\pm)}(\omega, r)=-\frac{ir(\nu_{(e\pm)} r\pr_{r_*} +\omega)}{q_r^{(e\pm)}}\bar{R}^{(e\pm)}(\omega, r),\\
&u_{(2)}^{(e\pm)}(\omega, r)=\frac{r(\pr_{r_*}-\omega\nu_{(e\pm)} r)}{q_r^{(e\pm)}}\bar{R}^{(e\pm)}(\omega, r).
\eea
\ee
we can write the mode function in near horizon limit as:
\begin{align}
u_{(1)}(\omega, r) &\xrightarrow{r \to r_h} \frac{\omega r_h}{q_h} (\nu r_h - i) \bar{R}_{\rm up}, \\
u_{(2)}(\omega, r) &\xrightarrow{r \to r_h} -\frac{\omega r_h}{q_h} (\nu r_h - i) \bar{R}_{\rm up}.
\end{align}
The vector field in null coordinates ($t=\frac{u+v}{2},r_*=\frac{v-u}{2}$) can be given as:
\begin{align}
    A_{u}&=\frac{u_{(1)}-u_{(2)}}{2r}Y_{\ell m}\,, & A_{v}&=\frac{u_{(1)}+u_{(2)}}{2r}Y_{\ell m}\,,\\
    A_{\theta}&=A_{\theta}\,,
    &A_{\varphi}&=A_{\varphi}.
\end{align}
The knowledge of radial solution allows us to write the mode function in near horizon limit as:
\begin{align}
u_{(1)}(\omega, r) &\xrightarrow{r \to r_h} \frac{\omega r_h}{q_h^{\pm}} (\nu_{\pm} r_h - i) \bar{R}^{(e\pm)}_{\rm up}, \\
u_{(2)}(\omega, r) &\xrightarrow{r \to r_h} -\frac{\omega r_h}{q_h^{\pm}} (\nu_{\pm} r_h - i) \bar{R}^{(e\pm)}_{\rm up}.
\end{align}
Accordingly, in the null coordinates the components of the vector field could be stated as follows:
Due to the structure of solution, $F_{uv}$ vanishes on the past horizon and $g^{uv}\to-\infty$ however the combined product stays well defined. The resulting object in the near horizon limit can be given as,
\be\label{gen-exp:Ftr}
\bea
&g^{uv}F_{vu}\\%&=F_{tr}=\pr_t A_r-\pr_r A_t\nt
&=i\bigg[\frac{\nu_{\pm} r V_{0}(r)}{q_r^{\pm}}  - \frac{\nu_{\pm}}{q_r^{\pm}} \partial_{r_*} - \frac{2 \omega \nu^2_{\pm} r}{q_r^{\pm}\,^2}\bigg]\bar{R}^{(e\pm)}_{\rm up}(r_*)Y_{\ell m}e^{-i\omega\,t},
\eea
\ee
where
\be
V_0(r)=\left(\frac{L}{r^2}+\mu^2+\frac{2\omega\nu}{q_r}\right).
\ee
Evaluating the inner product on the null boundary $d\Sigma^{u} = g^{uv}d\Sigma_v=-r_h^2 du d\Omega$ yields
\be
\bea
&(A^{{\rm (e\pm)}}_{\omega\,\ell m}, A^{{\rm (e\pm)}}_{\omega'\,\ell'm'})_{\mathcal{H}^-}\\
&= -i \int d\Sigma^u g^{\nu\alpha}\bigg[A^{(\pm)}_{\alpha\,\omega\,\ell m} F_{\nu u\,\omega'\,\ell'm'}^{(\pm)*}- F_{\nu u\,\omega\,\ell m}^{(\pm)} A_{\alpha\,\omega'\,\ell'm'}^{(\pm)*} \bigg]\\
&= i\int_{-\infty}^{\infty} du \int d\Omega \, r_h^2\left[ A_{u} F_{tr}^* - F_{tr} A^{\prime *}_u \right]\\
&+i \int_{-\infty}^{\infty} du \int d\Omega \, r_h^2 g^{AB}\left[A_{A} F_{B u}^{'*} - F_{B u} A_{A}^{'*} \right]\\
&=\frac{4\pi \omega r_h^2}{q_h^{\pm}}(\mu^2+ \nu^2_{\pm}L)|\mathcal{N}^{\rm(e\pm)\rm up}_{\omega \ell }|^2 \delta(\omega - \omega') \delta_{\ell\ell'} \delta_{mm'}.
\eea
\ee
To calculate the normalization constant for the up-modes, we first determine the asymptotic form of the Proca field near $\mathcal{I}^-$. At leading order in $r$, the first two components take the form
\begin{align}
A_v &\equiv \frac{c_v}{r} e^{-i\Omega_+ v-i\Omega_- u}Y_{\ell m},\\
A_u &\equiv \frac{c_u}{r} e^{-i\Omega_+ v-i\Omega_- u}Y_{\ell m},
\end{align}
where
\begin{align}
\Omega_+ \equiv \frac{\omega+k}{2},\qquad\Omega_- \equiv \frac{\omega-k}{2},
\end{align}
and
\begin{align}
c_v^\pm &= -\frac{\Omega_+}{\nu_\pm}\,\mathcal{N}_{\omega\ell}^{(e\pm)\,\rm in},&
c_u^\pm = \frac{\Omega_-}{\nu_\pm}\,\mathcal{N}_{\omega\ell}^{(e\pm)\,\rm in}.
\end{align}
These expressions automatically satisfy the Lorenz condition, $\nabla_\mu A^\mu =0,$ through the relation
\begin{align}
A_u = -\frac{\Omega_-}{\Omega_+}A_v.
\end{align}
At the same order, the electromagnetic field strength tensor simplifies to
\begin{align}
F_{uv}\approx -2i\Omega_- A_v,
\end{align}
which agrees with \eqref{gen-exp:Ftr} at $\mathcal{O}(r^{-1})$ upon using $V_0\approx \mu^2$. Evaluating the inner product on the null boundary with
\begin{align}
d\Sigma^v = g^{uv}d\Sigma_u= -r^2\,dv\,d\Omega,
\end{align}
we obtain
\be
\bea
&(A^{(\pm)\,\omega\ell m},A^{(\pm)\,\omega'\ell'm'})_{\mathcal{I}^-}\\
&=-i\int d\Sigma^u\, g^{\nu\alpha}\bigg[A^{(\pm)}_{\alpha\,\omega\ell m}F_{\nu v\,\omega'\ell'm'}^{(\pm)*}-F_{\nu v\,\omega\ell m}^{(\pm)}A_{\alpha\,\omega'\ell'm'}^{(\pm)*}\bigg]\\
&=-2i\int_{-\infty}^{\infty}dv\int d\Omega\,r^2\left(A_vF_{uv}^{\prime *}-F_{uv}A_v^{\prime *}\right)\\
&+i\int_{-\infty}^{\infty}dv\int d\Omega\,r^2 g^{CD}\left(A_CF_{Dv}^{\prime *}-F_{Dv}A_C^{\prime *}\right)\\
&=4\pi\left[4\Omega_-|c_v^\pm|^2+\Omega_+L\left|\mathcal{N}_{\omega\ell}^{(e\pm)\,\rm in}\right|^2\right]\delta(\Omega_+-\Omega_+')\delta_{\ell\ell'}\delta_{mm'},
\eea
\ee
where $d\Omega=\sin\theta\,d\theta\,d\varphi$. Now, using \cite{abramowitz1964handbook},
\be
\delta(\Omega_+-\Omega_+')=\frac{k}{|\Omega_+|}\delta(\omega-\omega'),
\ee
we find the following 
\be
\bea    
&(A^{{\rm (e\pm)\, in;}\,\omega\,\ell m}, A^{{\rm (e\pm)\, in;}\,\omega'\,\ell'm'})_{\mathcal{I}^-}\\
&~~~~= 4\pi k\left(\frac{\mu^2}{\nu^2_{\pm}} +L\right)|\mathcal{N}^{\rm (e\pm)\, in}_{\omega \ell }|^2 \delta(\omega - \omega') \delta_{\ell\ell'} \delta_{mm'}.\label{b13}
\eea
\ee
\begin{comment}
which then fixes the normalization as:
\begin{equation}
|\mathcal{N}_{\omega \ell }^{(e\pm)\,\rm in}| = \frac{|\nu_{\pm}|}{\sqrt{4\pi k(\mu^2+\nu^2_{\pm}L)}}.
\end{equation}
Notably in deriving \eqref{b13}, the asymptotic limit $r\to\infty$ was taken first. The angular contribution contains the factor $L(1-\frac{1}{q_r})$, making the $\mu\to0$ limit of the $\nu_-$ branch subtle since the limits do not commute:
\begin{align}
\lim_{r\to\infty}\lim_{\nu\to0}
\left(1-\frac{1}{1+\nu^2 r^2}\right)=0,
\end{align}
whereas
\begin{align}
\lim_{\nu\to0}\lim_{r\to\infty}
\left(1-\frac{1}{1+\nu^2 r^2}\right)=1.
\end{align}
\end{comment}
The only nontrivial case is the mixed inner product between the $e+$ and $e-$ modes. On $\mathcal{H}^-$, the inner product reduces to
\be
\bea
(A^+,A^-)_{\mathcal{H}^-}&=-i\int d\Sigma^u\,g^{\nu\alpha}\left[
A_\alpha^+ F_{\nu u}^{-*}-F_{\nu u}^+ A_\alpha^{-*}\right]\\
&=i\int_{-\infty}^{\infty}du\int d\Omega\, r_h^2\left(A_u^+F_{tr}^{-*}
-F_{tr}^+A_u^{-*}\right)\\
&+i\int_{-\infty}^{\infty}du\int d\Omega\, r_h^2g^{AB}\left(A_B^+F_{Au}^{-*}
-F_{Au}^+A_B^{-*}\right).
\eea
\ee
The expression involving the radial component in the above equation can be simplified and expressed as follows:
\begin{widetext}
\begin{multline}
A_u^+F_{tr}^{-*}-F_{tr}^+A_u^{-*}=\frac{\omega\mathcal{N}^{(e+)\,\rm up}_{\omega\ell}\mathcal{N}^{(e-)\,\rm up*}_{\omega\ell}}{q_h^+q_h^-}
\Bigg[(\nu_+-\nu_-)r_h\left(\mu^2+\frac{L}{r_h^2}\right)-i\left\{2\nu_+\nu_-r_h^2\left(\mu^2+\frac{L}{r_h^2}\right)+\omega(\nu_++\nu_-)\right\}\Bigg].
\end{multline}
\end{widetext}
The angular contribution follows from
\begin{align}
F_{Au}^{\pm}=\partial_AA_u^\pm-\partial_uA_A^\pm.
\end{align}
Separating the terms involving $\partial_u A_A^\pm$ from those involving
$\partial_A A_u^\pm$, one obtains
\begin{multline}
i\int_{-\infty}^{\infty}du\int d\Omega\,r_h^2g^{AB}\left(A_B^+F_{Au}^{-*}-F_{Au}^+A_B^{-*}\right)\\
=4\pi\omega L \mathcal{N}^{(e+)\,\rm up}_{\omega\ell}\mathcal{N}^{(e-)\,\rm up*}_{\omega\ell}\bigg[1-\frac{2-i(\nu_+-\nu_-)r_h}{
2(1+i\nu_-r_h)(1-i\nu_+r_h)}\bigg]\\
\times\delta(\omega-\omega')\delta_{\ell\ell'}\delta_{mm'}.
\end{multline}
The mixed inner product becomes,
\begin{widetext}
\begin{multline}
(A^{(e+)\,\rm up}_{\omega\ell m},A^{(e-)\, \rm up}_{\omega'\ell'm'})_{\mathcal{H}^-}=\frac{1}{2\sqrt{(\mu^2+\nu_+^2L)(\mu^2+\nu_-^2L)}}\Bigg[\frac{\left(2\nu_+\nu_- r_h^2+i r_h(\nu_+-\nu_-)\right)\left(\mu^2+\frac{L}{r_h^2}\right)+\omega(\nu_++\nu_-)}{\sqrt{q_h^+q_h^-}}\\
+\frac{2L\sqrt{q_h^+q_h^-}}{r_h^2}\left(1-\frac{2-i(\nu_+-\nu_-)r_h}{
2(1+i\nu_-r_h)(1-i\nu_+r_h)}\right)\Bigg]\delta(\omega-\omega')\delta_{\ell\ell'}\delta_{mm'}.
\end{multline}
\end{widetext}
Finally, using equation \eqref{eq22}, the above expression vanishes identically. Hence,
\begin{align}
(A^{(e+)\,\rm up}_{\omega\ell m},A^{(e-)\, \rm up}_{\omega'\ell'm'})_{\mathcal{H}^-}=0.
\end{align}
Similarly for in- modes
\be
(A^{(e+)\,\rm in}_{\omega\ell m},A^{(e-)\, \rm in}_{\omega'\ell'm'})_{\mathcal{I}^-}=\frac{\left(\mu^2+\nu_{+}\nu_{-}L\right)}{\sqrt{(\mu^2+\nu^2_{+}L)(\mu^2+\nu^2_{-}L)}}=0.
\ee
This concludes the analysis regarding the orthonormality of the FKKS ansatz.
%%%%%%%%%%%%%%%%%%%%%%%%%%%%%%%%%%%%%%%%%
\section{Derivation of Dirac Bracket}\label{append: diracbracket}
\begin{comment}
To contextualize the canonical formulation of the Proca field, it is necessary to review the fundamental shift in how time is treated in Hamiltonian formalism compared to the covariant Lagrangian framework. In the Lagrangian perspective, space and time are treated on equal footing; the fundamental object is, then, a field defined over the entire spacetime manifold, $A_\mu(t, x)$ with $x=(x_1,x_2,x_3)$.

The Hamiltonian formalism, however, relies on a $3+1$ decomposition of spacetime. The fundamental variables transition from being a spacetime function to spatial configurations, $A_\mu(x)$, defined on a fixed timelike hypersurface. These spatial configurations serve as the independent canonical coordinates of an infinite-dimensional phase space, while time $t$ is separated out as an external evolution parameter. Consequently, the phase space variables possess no explicit time dependence ($\partial_t A_\mu \equiv 0$). Their time evolution is strictly dynamical, generated entirely by Poisson brackets with the Hamiltonian.

With this distinction established,
\end{comment}
For singular Hamiltonian systems such as ours, the canonical momenta cannot be inverted uniquely to express all generalized velocities \cite{Henneaux:1994lbw, Brown:2022gha}. As a result, the system possesses constraints, and special care is required in defining the Hamiltonian evolution. Without introducing the appropriate Lagrange multipliers and the correct bracket structure, the equations of motion derived from the Hamiltonian formalism need not reproduce the Euler--Lagrange equations consistently. For this reason, one employs the Dirac bracket formalism to correctly handle the constraints. From the definition of the conjugate momentum given in the main text, \eqref{eq: conj_momentum}, we find that
\begin{equation}
    \Pi^0=0,\qquad\Pi^i=\sqrt{-g}\,F^{i0}.
\end{equation}
The vanishing of conjugate momenta is the primary constraint that the system possesses \cite{dirac2013lectures, Brown:2022gha},
\begin{equation}
    \Phi_1(x)\equiv \Pi^0(x)\approx0.
\end{equation}
The last weak equality helps one to proceed with the primary constraint and include it in the Hamiltonian density corresponding to the Proca action \eqref{eq: proca_action}, by the use of a Lagrange multiplier as,
\be
\bea
\mathcal{H}&=\frac{1}{2\sqrt{-g}}\,g_{ij}\Pi^i\Pi^j + \frac{\sqrt{-g}}{4}F_{ij}F^{ij} + \Pi^i\partial_i A_0\\
&~~~~+\frac12 \mu^2\sqrt{-g}   \left(-A_0^2 + g^{ij}A_iA_j\right)+\lambda\Pi^0.
\eea
\ee
To determine which physical degree of freedom is encoded with the Lagrange multiplier, we consider Hamilton's equation of $A_0$,
\be
\dot A_0 =\{A_0,H\}=\lambda.
\ee
Then, for the consistent evolution of the system, we require that the constraint does not evolve in time, i.e., $\dot\Phi_1\approx0$, thus we find the secondary constraint. To explicitly determine the expression associated with the secondary constraint, let us consider the following equation of motion
\begin{align}
\Phi_2=\dot{\Pi}^0(x)= \{\Pi^0(x),H\}.
%&= \int d^3y\,\Big(\{\Pi^0(x),\Pi^i(y)\partial_i A_0(y)\}+\{\Pi^0(x),-\tfrac12 m^2\sqrt{-g}A_0^2(y)\}\Big)\\
%&= \int d^3y\,\Big(-\Pi^i(y)\partial_i\delta^{(3)}(x-y) - m^2\sqrt{-g}A_0(y)\delta^{(3)}(x-y)\Big)\\
\end{align} 
The above Poisson bracket can be directly evaluated from the Hamiltonian by utilizing the fundamental Poisson brackets \eqref{eq: canonical_poisson_bracket}, and we get
\be
\Phi_2=\partial_i\Pi^i(x)-\mu^2\sqrt{-g}A_0(x).
\ee
%This could be derived from the perspective of EoM by considering $\nu=0$ component and using the definition of conjugate momenta as:
%\begin{equation}
%    \nabla_\mu F^{\mu 0} =\frac{1}{\sqrt{-g}}\partial_\mu(\sqrt{-g}F^{\mu 0}), \qquad F^{i0}=\frac{1}{\sqrt{-g}}\Pi^i,
%\end{equation}
%we obtain:
%\begin{equation}
%    \frac{1}{\sqrt{-g}}\partial_i(\Pi^i)-m^2 A^0=0.
%\end{equation}
%Thus we arrive at the secondary constraint:
%\begin{equation}
%    \Phi_2(x)\equiv \partial_i\Pi^i(x)-m^2\sqrt{-g(x)}\,A^0(x)\approx0.
%\end{equation}
%Incorporating the above expression back into the Hamiltonian via Lagrange multiplier and require $\dot\Phi_2=0$, which leads to 
To check whether the above form indeed leads to the secondary constraint, let us check its evolution as follows
\begin{widetext}
\be
\bea
\dot\Phi_2&=\{\partial_i\Pi^i(x)-\mu^2\sqrt{-g(t,x)}\,A^0(x),H\}+\pdv{\Phi_2}{t}\\
  %  &=\partial_i\underbrace{\{\Pi^i(x),H\}}_{\pdv{H}{A_i}}-m^2\sqrt{-g}\underbrace{\{\,A^0(x),H\}}_{\dot A^0}-m^2\pdv{\sqrt{-g}}{t}A^0\nt
&=\partial_i\left[\partial_j\!\left(\sqrt{-g}\,F^{ji}\right)  - \mu^2\sqrt{-g}\,g^{ij}A_j\right]-\mu^2\sqrt{-g}\dot{A^0}-\mu^2\partial_0(\sqrt{-g})A^0\\
&=-\mu^2\sqrt{-g}(\nabla_{i}A^{i}+\nabla_{0}A^0)=0.
\eea
\ee
\end{widetext}
It is interesting to realize that the Lorenz condition \eqref{eq: gauge.cond}, used in the last line, appears here as an integrability criterion, which fixes the Lagrange multiplier and prevents tertiary and higher-order constraints from appearing in the system. Here we have also treated $A_0$ as a fundamental phase space variable and that it evolves explicitly under the Poisson bracket with respect to the Hamiltonian $(\pr_t A_0=0)$.  

The fundamental equal–time Poisson brackets are as follows:
\begin{align}\label{eq: canonical_poisson_bracket}
 \{A_\mu(x),\Pi^\nu(y)\}=\delta_\mu^{\ \nu}\,\delta^{(3)}(x-y),\\
 \{A_\mu,A_\nu\}=\{\Pi^\mu,\Pi^\nu\}=0.    
\end{align}
As discussed earlier, above does not hold true for constraint $(\Pi^0=0)$. Therefore, at this stage one introduces the notion of weak equality and enforce that we do not use the primary constraint inside the Poisson bracket but only after evaluating the derivatives. This means, in particular, that it has nonzero Poisson brackets with the canonical variables. Since $A^0=g^{0\alpha}A_\alpha$, the fundamental equal time Poisson bracket for it becomes:
\begin{equation}
    \{A^0(x),\Pi^0(y)\}=g^{00}(x)\delta^{(3)}(x-y).
\end{equation}
Next, to determine the consistent Dirac brackets, we must compute the non-vanishing Poisson brackets among the constraints, which are given by:
\be
\bea
\{\Phi_1(x),\Phi_2(y)\}&=\{\Pi^0(x),-\mu^2\sqrt{-g(y)}A^0(y)\}\notag\\
&=-\mu^2\sqrt{-g(y)}\,g^{00}(y)\delta^{(3)}(x-y),
\eea
\ee
and similarly,
\begin{equation}
    \{\Phi_2(x),\Phi_1(y)\}=\mu^2\sqrt{-g(x)}\,g^{00}(x)\delta^{(3)}(x-y).
\end{equation}
Since all other constraint brackets vanish by virtue of the antisymmetry of the Poisson bracket, the constraint matrix takes the form:
\be
\bea
C_{ab}(x,y)&=\{\Phi_a(x),\Phi_b(y)\}\notag\\
&=\mu^2\sqrt{-g(x)}\,g^{00}(x)
\begin{pmatrix}
0 & -1\\
1 & 0
\end{pmatrix}
\delta^{(3)}(x-y).
\eea
\ee
The next step is to determine the inverse of this constraint matrix by solving the following equation:
\begin{equation}
    \int d^3z\,(C^{-1})^{ac}(x,z)C_{cb}(z,y)=\delta^a_{\ b}\delta^{(3)}(x-y).
\end{equation}
This gives us the inverse of constraint matrix as follows:
\begin{equation}
    (C^{-1})^{ab}(x,y)=
\frac{1}{\mu^2\sqrt{-g(x)}\,g^{00}(x)}
\begin{pmatrix}
0 & -1\\
1 & 0
\end{pmatrix}
\delta^{(3)}(x-y).
\end{equation}
With this structure in place, we now evaluate the Dirac bracket given by,
\be
\bea
\{F,G\}_D &=\{F,G\}\\
&-\int d^3u\,d^3v\,\{F,\Phi_a(u)\}(C^{-1})^{ab}(u,v)\{\Phi_b(v),G\}.
\eea
\ee
Notably, not all of the fundamental Poisson brackets are modified by the Dirac bracket construction. Therefore, here we only outline the ones that receive non-trivial corrections. The first of which can be found from:
\begin{widetext}
    \begin{align}
    &\{A^0(x),\Pi_i(y)\}_D=\underbrace{\{A^0(x),\Pi_i(y)\}}_{=0}-\int d^3u\,d^3v\,\{A^0(x),\Phi_a(u)\}(C^{-1})^{ab}(u,v)\{\Phi_b(v),\Pi_i(y)\}.\label{eq: A0Pi}
\end{align}
\end{widetext}
To evaluate this, we need the following Poisson brackets of constraints with field and conjugate momenta
\begin{align}
\{A^0(x),\Phi_1(u)\}
&=\{A^0(x),\Pi^0(u)\}
=g^{00}(x)\delta^{(3)}(x-u),\\
\{A^0(x),\Phi_2(u)\}&=0,\\
\{\Phi_2(v),\Pi_i(y)\}
&=\{\partial_k\Pi^k(v),\Pi_i(y)\}\notag\\
&\quad-\mu^2\sqrt{-g(v)}\{A^0(v),\Pi^i(y)\}
=0\\
\{\Phi_1(v),\Pi_i(y)\}&=0.
\end{align}
Using these, \eqref{eq: A0Pi} yields,
\begin{equation}
    \{A^0(x),\Pi_i(y)\}_D=0.
\end{equation}
The next non-trivial Dirac bracket that differs from their Poisson bracket counterpart can be given as:
\begin{widetext}
    \begin{align}
     \{A^0(x),A_i(y)\}_D=\underbrace{\{A^0(x),A_i(y)\}}_{=0}&-\int d^3u\,d^3v\,\{A^0(x),\Phi_a(u)\}(C^{-1})^{ab}(u,v)\{\Phi_b(v),A_i(y)\}.\label{eq: A0Ai}
\end{align}
\end{widetext}
The relevant Poisson brackets for evaluating it can be given as
\begin{align}
\{A^0(x),\Phi_1(u)\}&=\{A^0(x),\Pi^0(u)\}=g^{00}(x)\delta^{(3)}(x-u),\\
\{A^0(x),\Phi_2(u)\}&=0,\\
\{\Phi_2(v),A_i(y)\}&=\{\partial_k\Pi^k(v),A_i(y)\}=-\partial_i\delta^{(3)}(v-y),\\
\{\Phi_1(v),A_i(y)\}&=0.
\end{align}
Thus, \eqref{eq: A0Ai} becomes
\begin{align}
\{A^0(x),A_i(y)\}_D%&=-\int d^3u\,d^3v\;g^{00}(x)\delta(x-u)\frac{-1}{m^2\sqrt{-g(u)}g^{00}(u)}\delta(u-v)\left(-\partial_i\delta(v-y)\right)\\
&=-\frac{1}{\mu^2\sqrt{-g(x)}}\,\partial_{x^i}\delta^{(3)}(x-y).
\end{align}
Rest of the Dirac Bracket are identitcal to their Poisson bracket counterpart. Therefore, collectively we can state them here as:
\begin{align}
 &\{A_i(x),\Pi^j(y)\}_D=\delta_{i}^{j}\,\delta^{(3)}(x-y),\\
 &\{A_i,A_j\}_D=\{\Pi^i,\Pi^j\}_D=0,\\
 &\{A^0(x),\Pi_i(y)\}_D=0,\\
 &\{A^0(x),A_i(y)\}_D=
-\frac{1}{\mu^2\sqrt{-g(x)}}\,\partial_{x^i}\delta^{(3)}(x-y),
\end{align}
where $x=(x_1,x_2,x_3)$ and $y=(y_1,y_2,y_3)$. In order to quantize the system we promote $\{\, , \,\}_D \;\longrightarrow\; \frac{1}{i}[\, , \,]$. Then the analysis from \Cref{sec: quantization_and_vacua} follows.
\section{Two-Point Correlators and Asymptotic Structure}\label{append: twopointasymptotic}
Using the mode expansion in \eqref{mode-decom}, the quantized Proca field can be written as
\be
\bea
\hat{A}_{\mu}&= \int_{0}^{\mu}\,d\omega\sum_{\lambda\,\ell m}\hat a_{\omega\,\ell m}^{(\lambda)} A_{\mu\,\omega \ell m}^{(\lambda)\,\rm up}(x)+a_{\omega\,\ell m}^{(\lambda)\dagger} A_{\mu\,\omega \ell m}^{(\lambda)\,\rm up*}(x)\\
&+\sum_{n}\int_{\mu}^{\infty}\,d\omega\sum_{\lambda\,\ell m}\hat a_{\omega\,\ell m}^{(\lambda)\,n} A_{\mu\,\omega \ell m}^{(\lambda)\,n}(x)+a_{\omega\,\ell m}^{(\lambda)\,n\dagger} A_{\mu\,\omega \ell m}^{(\lambda)\,n*}(x).
\eea
\ee
The evaluation of the two-point function reduces to computing $A^{(\lambda)}_{\mu\,\omega \ell m}(x) A^{(\lambda)*}_{\nu\,\omega \ell m}(x')$ using the mode function decomposition given in \eqref{eq: proca_decomp} and subsequently sum over $\ell,m$. It is convenient to first perform the sum over $m$. For fixed polarization $\lambda$ and angular momentum $\ell$, one obtains
\begin{widetext}
\be
\lim_{x\to x'}\sum_{m=-\ell}^{\ell} A^{(\lambda)}_{\mu\,\omega \ell m}(x) A^{(\lambda)*}_{\nu\,\omega \ell m}(x') = \frac{2\ell+1}{4\pi r^2} 
\begin{pmatrix} 
|u_1^{(\lambda)}|^2 & \frac{ u_1^{(\lambda)} u_2^{\lambda\star}}{f} & 0 & 0 \\ 
\frac{u_1^{\lambda\star} u_2^{(\lambda)}}{f} & \frac{|u_2^{(\lambda)}|^2}{f^2} & 0 & 0 \\ 
0 & 0 &\frac{r^2(|u_3^{(\lambda)}|^2 + |u_4^{(\lambda)}|^2)}{2L}  & 0 \\ 
0 & 0 & 0 & \frac{r^2(|u_3^{(\lambda)}|^2 + |u_4^{(\lambda)}|^2)}{2L} \sin^2\theta 
\end{pmatrix},
\ee
where we have used the following relations,
\begin{equation}
    \sum_{m=-\ell}^{\ell}|Y_{\ell m}|^2 =\frac{2\ell+1}{4\pi}; \qquad
\sum_{m} \partial_A Y_{\ell m} \partial_B \bar{Y}_{\ell m} = \frac{L(2\ell+1)}{8\pi}\rm{diag}(1,\sin^2\theta);\qquad \sum_{m} Y_{\ell m} \partial_A \bar{Y}_{\ell m}=0.
\end{equation}
\end{widetext}
The resulting correlator is stationary, with no explicit dependence on the Schwarzschild time coordinate. Furthermore, the angular sector becomes independent of the radial falloff apart from the overall prefactor.

Following $\Delta G_{\mu\nu}^{\Psi-B}=\bra{\Psi}\hat A_\mu \hat A_\nu \ket{\Psi}-\bra{0_B}\hat A_\mu \hat A_\nu \ket{0_B}$ and utilising \eqref{GB}--\eqref{GH} for the Unruh vacuum $(\ket{\Psi}=\ket{0_U})$, one can identify that the finite contribution to the two-point function is given by
\begin{widetext}
\begin{equation}\label{GUB}
\Delta G_{\mu\nu}^{(U-B)}(r) = \sum_{\lambda, \ell} \frac{2\ell+1}{4\pi r^2} \int_{0}^{\infty} d\omega \left[ \frac{2}{e^{\frac{\pi\omega}{\kappa}}-1}\right] \begin{pmatrix} 
|u_1^{\rm up}|^2 & \frac{ u_1^{\rm up} u^{\rm up*}_2}{f} & 0 & 0 \\ 
\frac{u^{\rm up*}_1 u_2^{\rm up}}{f} & \frac{|u_2^{\rm up}|^2}{f^2} & 0 & 0 \\ 
0 & 0 &\frac{r^2( |u_3^{\rm up}|^2 + |u_4^{\rm up}|^2)}{2L}  & 0 \\ 
0 & 0 & 0 & \frac{r^2( |u_3^{\rm up}|^2 + |u_4^{\rm up}|^2)}{2L} \sin^2\theta
\end{pmatrix}.
\end{equation}
Here, and in the expression below, the polarization index, $\lambda$, with the radial functions should be assumed. Similarly, for the Hartle--Hawking vacuum $(\ket{\Psi}=\ket{0_{HH}})$ one finds
\begin{equation}
\Delta G_{\mu\nu}^{(HH-B)}(r) = \sum_{\substack{\rm n\in\{\rm in, up\}\\ \lambda,\ell}} \frac{2\ell+1}{2\pi r^2} \int_{0}^{\infty} d\omega\frac{1}{e^{\frac{\pi\omega}{\kappa}}-1} 
\begin{pmatrix} 
|u_1^{\rm n}|^2 & \frac{ u_1^{\rm n} u^{\rm n*}_2}{f} & 0 & 0 \\ 
\frac{u^{\rm n*}_1 u_2^{\rm n}}{f} & \frac{|u_2^{\rm n}|^2}{f^2} & 0 & 0 \\ 
0 & 0 &\frac{r^2( |u_3^{\rm n}|^2 + |u_4^{\rm n}|^2)}{2L}  & 0 \\ 
0 & 0 & 0 & \frac{r^2( |u_3^{\rm n}|^2 + |u_4^{\rm n}|^2)}{2L} \sin^2\theta
\end{pmatrix}\label{GHH}.
\end{equation}
\end{widetext}
To understand the behavior of these correlators, it is useful to examine the asymptotic form of the mode functions.

For the even-parity up-modes near the future horizon $(\mathcal H^+)$, together with \eqref{Rup:BD}, the radial amplitudes behave as
\begin{comment}
\begin{align}
    &u_{(1)}(\omega, r)=-\frac{ir(\nu r\partial_{r_*} +\omega)}{q_r}\bar{R}(\omega, r)\\
    &u_{(2)}(\omega, r)=\frac{r(\partial_{r_*}-\omega\nu r)}{q_r}\bar{R}(\omega, r).
\end{align}
\end{comment}
\begin{align}
u_{(1)}^{(e\pm)}(\omega, r) &\xrightarrow{r \to r_h} -\mathcal{N}^{\rm (e\pm)\,\rm up}_{\omega \ell }\frac{\omega r_h}{q_h^{\pm}} (\nu_{\pm} r_h + i)  \mathcal{R}^{(e\pm)\,\text{up}}_{\omega \ell} e^{-i \omega r_*}, \\
u_{(2)}^{(e\pm)}(\omega, r) &\xrightarrow{r \to r_h} -\mathcal{N}^{\rm (e\pm)\,\rm up}_{\omega \ell }\frac{\omega r_h}{q_h^{\pm}} (\nu_{\pm} r_h + i)  \mathcal{R}^{(e\pm)\,\text{up}}_{\omega \ell} e^{-i \omega r_*},\\
u_{(3)}^{(e\pm)}(\omega,r) &\xrightarrow{r \to r_h} \mathcal{N}^{\rm (e\pm)\,\rm up}_{\omega \ell }L  \mathcal{R}^{(e\pm)\,\text{up}}_{\omega \ell} e^{-i \omega r_*},\\
u_{(4)}^{(e\pm)}(\omega,r) &\xrightarrow{r \to r_h} 0.
\end{align}
For the odd parity sector, near the future horizon $(\mathcal H^+)$, we get
\begin{align}
u_{(1)}^{({\rm o})}(\omega, r) &\xrightarrow{r \to r_h} 0,\\
u_{(2)}^{({\rm o})}(\omega, r) &\xrightarrow{r \to r_h} 0,\\
u_{(3)}^{({\rm o})}(\omega,r) &\xrightarrow{r \to r_h} 0,\\
u_{(4)}^{({\rm o})}(\omega,r) &\xrightarrow{r \to r_h} \mathcal{N}^{\rm ({\rm o})\,\rm up}_{\omega \ell }\mathcal{R}^{({\rm o})\,\text{up}}_{\omega \ell } e^{-i \omega r_*}.
\end{align}
At future null infinity $(\mathcal I^+)$, the even-parity sector behaves as
\begin{align}
u_{(1)}^{(e\pm)}(\omega, r) &\xrightarrow{r \to \infty} \frac{k}{\nu}\mathcal{N}^{\rm (e\pm)\,\rm up}_{\omega \ell } \mathcal{T}^{\lambda\,\text{up}}_{\omega \ell} e^{i k r_*}, \\
u_{(2)}^{(e\pm)}(\omega, r) &\xrightarrow{r \to \infty} -\frac{\omega}{\nu}\mathcal{N}^{\rm (e\pm)\,\rm up}_{\omega \ell }\mathcal{T}^{\lambda\,\text{up}}_{\omega \ell} e^{i k r_*}\\
u_{(3)}^{(e\pm)}(\omega,r) &\xrightarrow{r \to \infty} \mathcal{N}^{\rm (e\pm)\,\rm up}_{\omega \ell }L  \mathcal{T}^{(e\pm)\,\text{up}}_{\omega\ell} e^{ik r_*},\\
u_{(4)}^{(e\pm)}(\omega,r) &\xrightarrow{r \to \infty} 0.
\end{align}
For the odd parity sector, at future null infinity $(\mathcal I^+)$, we have
\begin{align}
u_{(1)}^{({\rm o})}(\omega, r) &\xrightarrow{r \to r_h} 0,\\
u_{(2)}^{({\rm o})}(\omega, r) &\xrightarrow{r \to r_h} 0,\\
u_{(3)}^{({\rm o})}(\omega,r) &\xrightarrow{r \to r_h} 0,\\
u_{(4)}^{({\rm o})}(\omega,r) &\xrightarrow{r \to r_h} \mathcal{N}^{\rm ({\rm o})\,\rm up}_{\omega \ell }\mathcal{T}^{({\rm o})\,\text{up}}_{\omega \ell } e^{i k r_*}.
\end{align}
The corresponding expressions for the in-modes follow from the replacements
\begin{align}
\mathcal{N}^{(\lambda)\,\rm up}_{\omega \ell }&\;\to\;\mathcal{N}^{(\lambda)\,\rm in}_{\omega \ell },&\mathcal{T}^{(\lambda)\,\text{up}}_{\omega \ell } &\;\to\;
\mathcal{R}^{(\lambda)\,\text{in}}_{\omega \ell },
\end{align}
with $\lambda=({\rm e\pm,o})$. 

\bibliography{Bibliography}
\end{document}